\documentclass[fleqn,usenatbib]{mnras}

\usepackage{newtxtext}
\usepackage[T1]{fontenc}
\usepackage{ae,aecompl}

\usepackage{graphicx}	% Including figure files
\usepackage{amsmath}	% Advanced maths commands
\usepackage{amssymb}	% Extra maths symbols
\usepackage{multirow}

\title[AstroVaDEr]{AstroVaDEr: Astronomical Variational Deep Embedder for Unsupervised Morphological Classification of Galaxies and Synthetic Image Generation}

\author[A. Spindler et. al.]{
Ashley Spindler\thanks{E-mail: a.spindler@herts.ac.uk},
James E. Geach, and Michael J. Smith
\\
Centre for Astrophysics Research, Department of Physics, Astronomy \& Mathematics, University of Hertfordshire, College Lane, Hatfield, AL10 9AB\\
Centre of Data Innovation Research, Department of Physics, Astronomy \& Mathematics, University of Hertfordshire, College Lane, Hatfield, AL10 9AB}
\date{Accepted XXX. Received YYY; in original form ZZZ}

\pubyear{2020}

\begin{document}
\label{firstpage}
%\pagerange{\pageref{firstpage}--\pageref{lastpage}}
\maketitle

% Abstract of the paper
\begin{abstract}
We present AstroVaDEr, a variational autoencoder designed to perform unsupervised clustering and synthetic image generation using astronomical imaging catalogues. The model is a convolutional neural network that learns to embed images into a low dimensional latent space, and simultaneously optimises a Gaussian Mixture Model (GMM) on the embedded vectors to cluster the training data. By utilising variational inference, we are able to use the learned GMM as a statistical prior on the latent space to facilitate random sampling and generation of synthetic images. We demonstrate AstroVaDEr's capabilities by training it on gray-scaled \textit{gri} images from the Sloan Digital Sky Survey, using a sample of galaxies that are classified by Galaxy Zoo 2. An unsupervised clustering model is found which separates galaxies based on learned morphological features such as axis ratio, surface brightness profile, orientation and the presence of companions. We use the learned mixture model to generate synthetic images of galaxies based on the morphological profiles of the Gaussian components. AstroVaDEr succeeds in producing a morphological classification scheme from unlabelled data, but unexpectedly places high importance on the presence of companion objects---demonstrating the importance of human interpretation. The network is scalable and flexible,  allowing for larger datasets to be classified, or different kinds of imaging data. We also demonstrate the generative properties of the model, which allow for realistic synthetic images of galaxies to be sampled from the learned classification scheme. These can be used to create synthetic image catalogs or to perform image processing tasks such as deblending.
\end{abstract}

% Select between one and six entries from the list of approved keywords.
% Don't make up new ones.

%%%%%%%%%%%%%%%%%%%%%%%%%%%%%%%%%%%%%%%%%%%%%%%%%%

%%%%%%%%%%%%%%%%% BODY OF PAPER %%%%%%%%%%%%%%%%%%

\section{Introduction}

Over the past hundred years, extragalactic astronomy has seen a continual evolution in the methods used to collect, process and analyse observational data. From photographic plates to electronic detectors, and from human computers to high-performance computer algorithms, the advancement of data acquisition and analysis techniques are not always in lock-step. Amongst the many data challenges that need to be solved, a particular problem arises when (human) visual classification of individual objects is the state-of-the-art methodology. Classifying the morphologies of galaxies is one such task, and until now it has been an achievable goal for teams of expert classifiers or citizen scientists to visually inspect every object within a survey. However, even at the best classification rates for crowd sourced projects, it would take many years to collect enough classifications for current generation surveys (e.g. DES \citep{doi:10.1142/S0217751X05025917}, DECaLS \citep{2019AJ....157..168D}, Hyper-Supreme Cam \citep{10.1093/pasj/psx066}), let alone the next-generation wide and deep surveys due to come on line within the next decade \citep{2020MNRAS.491.1554W}.

Addressing the challenge of producing physically and semantically meaningful morphological labels at the scales required for the Legacy Survey of Space and Time to be conducted with the Vera Rubin Observatory \citep{Ivezic:2008fe}, or images from \textit{Euclid}, invariably drives attention towards machine learning as a solution. In particular, deep neural networks have been demonstrated to great effect at the task of classifying certain morphological characteristics within galaxies \citep{LeCun.DeepLearning}. Until recently, most attention within the astronomical research community has been towards supervised machine learning techniques, where a labelled set of images is used to train a model to predict the class labels of new, unseen inputs. Artifical Neural Networks (ANNs) for galaxy classification have been in use in astrophysics for at least 25 years \citep[e.g.][]{1990Natur.348..221A, 1992ApJ...390L..41L, 1992MNRAS.259P...8S, 1993PASP..105.1354O, 1992AJ....103..318O, 1995Sci...267..859L}, and the recent growth in the power of neural networks has seen an upsurge in their application in astronomy. For example \cite{2015MNRAS.450.1441D} finished first place in the \textit{Galaxy Challenge} by using a rotationally-invariant convolutional neural network (CNN) to predict the voting fractions of galaxies that were classified by Galaxy Zoo 2. More recent approaches, such as \cite{2020MNRAS.491.1554W}, show the potential for supervised learning combined with crowd sourced labelling to actively improve and inform the neural network classifications. Supervised learning has also been demonstrated in the discovery of strong gravitational lenses, which is essentially a galaxy morphology classification problem \citep{2019MNRAS.484.3879P, 2020arXiv200402715L, 2019A&A...625A.119M, 2019ApJ...877...58A, 2018MNRAS.473.3895L}.

While supervised methods show some promise, they still face the limiting challenge of generating sufficient labelled images to form a cohesive training set that has imaging quality matching future surveys. However, recent work has shown that unsupervised learning methods also perform well at visually classifying astronomical objects. \cite{2020MNRAS.494.3750C} used a Convolutional Autoencoder (CAE), paired with Bayesian Gaussian Mixture Models (GMMs), to successfully construct a classifier for strong gravitational lenses. \cite{2019PASP..131j8011R} also exploited CAEs, combining the feature extraction capability of the autoencoder with a Self-Organised Map and {\it k}-means clustering to identify different classes of radio galaxy morphology. Using CAEs as a starting point provides many advantages, while retaining the image recognition power of supervised CNNs. The main benefit, of course, is in removing the need for a large volume of labelled training data.

Autoencoders work by learning how to transform data into a low-dimensional representation (sometimes called the latent space, embedding or encoded representation), and back again. This is most commonly achieved with a pair of neural networks, one to encode and one to decode the data \citep{37f2b6bee745402aa4e4d124d33be0e0, Boulard.Kamp, 10.5555/2987189.2987190}. The methods discussed above rely on separating out the dimensionality reduction and clustering tasks, however it is possible to combine the tasks by using so-called `Deep clustering' techniques \citep{2015arXiv151106335X,2016arXiv161102648D}. By training the embedding and clustering processes simultaneously, the learned latent space is encouraged to take on a clustered distribution while the clustering parameters evolve to follow the latent space. Deep clustering techniques have been shown to produce higher accuracy clustering scores on standard data sets than independently optimised solutions, and promise to produce more interpretable clusters \citep{2019arXiv191106623R, 2016arXiv161105148J, 2020arXiv200508047C}. 

Deep learning is not the only form of unsupervised methodology which has been employed in astronomy. \cite{2018MNRAS.473.1108H} and \cite{2020MNRAS.491.1408M} demonstrate the ability of using a Growing Neural Gas and Hierarchical Clustering algorithm, which populates a learned model with low dimensional neurons that act as representations of different galaxy properties. Instead of learning a feature variable from scratch, as a deep CNN would, \cite{2018MNRAS.473.1108H} and \cite{2020MNRAS.491.1408M} use Fourier transformed image patches to encode morphological and spectral information, training the model to group together similar patches into objects, and similar objects into morphological clusters. \cite{2020arXiv200406734U} use principal component analysis of {\it Hubble Space Telescope}-CANDELS images to demonstrate how the morphological features of galaxies can be represented by `eigengalaxies' which describe different components of an underlying morphological manifold. We have also seen implementations in time-domain astronomy, such as in \cite{2020MNRAS.493..713A}, which uses a Dirchelet-Process GMM to identify different classes of pulsars from their periods and period derivatives. Finally, the prediction of redshifts from photometric data has also been studied with unsupervised learning, such as in \cite{2012MNRAS.419.2633G}, \cite{2018arXiv180509905S}, and \cite{2018A&A...609A.111D}.

There have been promising developments in recent years in improving the ability of autoencoders (AEs) to perform image classification, and among those is the integration of Variational Inference \citep{2016arXiv160100670B} into the encoding process. Variational inference is a field of statistics concerned with finding approximations of the posterior distributions in Bayesian Models. \cite{2013arXiv1312.6114K} demonstrated how a variational inference model could be approximated by using the autoencoder framework, which has led to a wide field of research into the applications of Variational Autoencoders (VAE). Like CAEs, a VAE encodes a sample of data into a low-dimensional space, but it differs in the sense that a statistical prior is used to `condition' the encoded space to take a certain shape. The most common prior in a VAE is a unit Gaussian, however other models have been designed that allow for deep clustering. Variational Deep Embedding \citep[VaDE,][]{2016arXiv161105148J} represents one of the state-of-the-art approaches for deep, unsupervised clustering, and works by imposing a Gaussian Mixture prior upon the learned latent space.

An important distinction between a VAE and a traditional AE is that a VAE is in fact a generative network. We will discuss the generative process in Section \ref{sec.methods}, but suffice to say that the statistical prior can be used to generate synthetic images. This is analogous to another generative network that has seen some popularity in astronomy: Generative Adversarial Networks (GANs) \citep{10.5555/2969033.2969125, 2016arXiv160505396R, 2019MNRAS.490.4985S, 2017MNRAS.467L.110S}. A VAE differs from a GAN in that the latter generally comprises two neural networks that are attempting to fool each other. One network (the generator) in the GAN is attempting to produce new data which matches the feature distribution of the training set, while the second network (the discriminator) attempts to guess if the new data is real or fake. The two networks compete, with the generator learning to make better (more realistic) synthetic data, and the discriminator getting better at distinguishing generated outputs from the real thing. 

Variational inference has already been explored in the context of extragalactic astronomy. Most applicable to this work is that of \cite{Regier2015ADG}, who demonstrate the fundamentals of using a VAE to embed morphological characteristics of galaxies. \cite{Ravanbakhsh2017EnablingDE} also implement a VAE, but do so using a conditional scheme that leverages known properties about the training images to guide the learning process. The conditional VAE (C-VAE) is shown alongside a conditional GAN network, and the authors show that the C-VAE produced more consistent results, and that by adding noise profiles associated with the original dataset they could produce well-realised synthetic images. While not a full Gaussian mixture, \cite{2019arXiv191014056S} successfully employ a Cascade-VAE with a double peaked Gaussian prior to perform star-galaxy separation. To our knowledge, the use of a Gaussian Mixture prior, as in VaDE, for galaxy morphological classification and image generation has not yet been demonstrated.

In this paper, we introduce AstroVaDEr (Astronomical Variational Deep Embedder), an implementation of the VaDE architecture which leverages numerous recent improvements to the variational deep clustering (VDC) paradigm. Here we demonstrate AstroVaDEr's capability as an unsupervised classifier for galaxy morphology, and show how its variational inference properties allow the network to be employed as a generative network. We perform training on, and comparisons with, galaxies from the Galaxy Zoo 2 \citep{2013MNRAS.435.2835W, 2015ApJS..221....8H} dataset to benchmark the network, but we also demonstrate some of the flexible development choices that allow AstroVaDEr to be adapted to other surveys and image classification problems.

In Section \ref{sec.Data} we describe the training, validation and testing sets we selected for this work, along with the image pre-processing that was performed. Section \ref{sec.methods} describes the theoretical background that informed the model and the chosen architecture, followed by details on hyperparameter selection and model training in Section \ref{sec:training}. Our results are presented in Section \ref{sec.results}, which includes a demonstration of the image reconstructions achieved through training, the unsupervised clustering results, a comparison with Galaxy Zoo 2 voting fractions and synthetic image generation properties. We discuss future improvements and our conclusions in Sections \ref{sec:future_work} and \ref{sec:conclusions}, respectively.

\section{Data}
\label{sec.Data}
In order to test the ability of VaDE to produce cluster assignments that are representative of the real underlying distribution of galaxies, we require a large dataset of labelled images of galaxies. For this purpose, we use the Galaxy Zoo 2 (GZ2) dataset, as described in \cite{2013MNRAS.435.2835W}. The images classified by Galaxy Zoo are taken from Data Release 7 of the Sloan Digital Sky Survey (SDSS), and make up a magnitude limited sample with $M_\text{r} < -17$\,mag. We extract our images from the SDSS \texttt{imgcutout} service, initially extracting \texttt{gri} colour images at $192\times192$ pixels, scaled to $0.2\times R90$ arcsec per pixel \footnote{$R90$ refers to the Petrosian radius containing $90\%$ of the galaxy's light.}. These images are cropped smaller than those used in the citizen science project to lessen the effect on the network of nearby structure unassociated with the target galaxy.

To help prevent over-fitting and to augment the data set, images are randomly flipped on their horizontal and vertical axes each time they are fed into the network. Mainly, this is to try and prevent the network learning some rare features in specific locations in the images, but the nature of autoencoders brings into question the validity of other image augmentation. For example, while CNNs can be made rotationally invariant \citep{2015MNRAS.450.1441D}, this invariance does not necessarily transfer to the fully-connected layers used within the latent embedding. Performing random rotations (which is a common augmentation technique in galaxy morphology studies) does not prevent the network from encoding rotational features. The network needs to be able to reproduce the image regardless of its observed orientation. Random rotations may improve disentanglement of rotation as a feature (i.e.\ the network would use less of the encoded space to control rotation), but may hamper the disentanglement of other features. Finally, there is a scalability consideration, as performing rotations on each input can prove costly in terms of processing time.

After the random transformations, we convert the image to grey scale by averaging the three colours bands. Training on grey scale images allows us to ensure that the network is learning strictly on the basis of morphology, as opposed to learning the colour dependence of different morphological types\footnote{An interesting alternative may be to train using CMYK format images, with a regulariser within the clustering model to emphasise the features in the luminance band over the colour bands.}. Finally, we downscale the images to $128\times128$ pixels to improve training speed.

Galaxy Zoo 2 provides labels for each galaxy in the sample in the form of vote counts for answers to a series of questions about a galaxy's morphology. These questions cover whether the galaxy is smooth or featured, and further covers topics such as the presence of bars and rings and the size of the bulge. We do not use the votes during training, but we will analyse our clustering results with respect to the raw and redshift-debiased vote fractions for a variety of morphological properties. Quantitatively, this will allow us to compare the properties the network chooses are most important versus the citizen scientists.

We select galaxies from the Galaxy Zoo 2 catalog that have more than 36 votes for the first question (smooth / features / star / artifact). We then split the sample into a training set, validation set and a test set. The training set uses a random selection of roughly 80\% of the catalog, we round to 100 objects to streamline mini-batch processing, with 159,600 galaxies in total. The validation set is a small selection of 5000 objects which is used to periodically test the network during training, but does not influence the learning process. The final test set consists of 41,100 objects. These are not seen during the training process by the network, but are used to perform analysis on the clustering model.

\section{Methods}
\label{sec.methods}

We combine several machine learning techniques to develop a network architecture that is flexible and powerful. Our hope is that this network is not simply tuned to the specific task of classifying galaxy morphology, but rather that the base architecture can be easily modified and optimised for a variety of tasks. AstroVaDEr is a combination of several staples of the `machine learning in astronomy' community, as well as more recent and modern algorithms tailored towards unsupervised learning. Our model builds on previous work within and without the field of astronomy. Firstly, our CNN is based on the work of \cite{2020MNRAS.491.1554W} (W20) in classifying galaxy morphology for Galaxy Zoo images. Secondly, we implement the VaDE technique from \cite{2016arXiv161105148J} (J16), with recent improvements to the algorithm discussed in \cite{2020arXiv200508047C} (C20) for `Simple, Scalable, and Stable Variational Deep Clustering' (s3VDC). In this section, we shall discuss the frameworks underlying the AstroVaDEr architecture and then present the specific model used in our galaxy morphology experiments.

\subsection{Autoencoders}

The foundational concept behind AstroVaDEr as an unsupervised clustering algorithm is the autoencoder. In essence, an autoencoder is a neural network that takes in some $N$-dimensional input, $\boldsymbol{x}$, reduces that input to a lower $n$-dimensional space (for $n\ll N$), and then reconstructs an output, $\boldsymbol{\hat{x}}$ which is compared with the input. An autoencoder optimises the weights and biases of neurons, typically in the form of fully connected and convolutional layers, by attempting to reduce the reconstruction loss between $\boldsymbol{x}$ and $\boldsymbol{\hat{x}}$. Practically, an autoencoder can be designed with two components, an encoder ($E_{\phi}$) and a decoder ($D_{\theta}$), such that:

\begin{equation}
    E_{\phi}(\boldsymbol{x}) = \boldsymbol{z}
\end{equation}

\begin{equation}
    D_{\theta}(\boldsymbol{z}) = \boldsymbol{\hat{x}}
\end{equation}

\noindent where $\phi$ and $\theta$ are the values of the weights and biases of the layers in the encoder and decoder, respectively. The latent variable, $\boldsymbol{z}$ is the encoded representation of the input. Autoencoders prove to be a reliable form of dimensionality reduction, which allows for unsupervised clustering algorithms to be used on data sets that would otherwise be too complex. For example, \cite{2020MNRAS.494.3750C} employ an autoencoder with convolutional layers to learn a latent representation from a data set of simulated gravitational lens imaging, which is then used to perform Bayesian Mixture Modelling to build a clustering and classification scheme for determining if a given image contains a gravitational lens. Note that there exist applications of autoencoders that embed the clustering process within the network itself. Deep Embedded Clustering optimises both the autoencoder parameters and the parameters of a unsupervised clustering or mixture model simultaneously, which results in increase classification accuracy.

For AstroVaDEr, autoencoders pose one crucial problem. Because we do not impose any structure on the latent space, it is inherently not a generative model. We do not know what the underlying statistical distribution is in the latent space, and as such we cannot create a random sample from it that could be fed into the decoder network to create synthetic images. However, it is possible to construct an autoencoder that has this property, by using the technique of variational inference.

\subsection{Variational Autoencoders}

Variational inference is a field of statistics concerned with finding approximations of the posterior distributions in Bayesian Models. \cite{2013arXiv1312.6114K} developed a framework wherein an autoencoder can be constructed with knowledge of a prior distribution on the latent space, such that in the process of optimising the neural network we approximate the posterior distribution. In this scheme, the encoder becomes an inference network that approximates the distribution $q_{\phi}(\boldsymbol{z} | \boldsymbol{x})$ by learning to map $\boldsymbol{x}$ to $\boldsymbol{z}$. The decoder is then a generative network, approximated the distribution $p_{\theta}(\hat{\boldsymbol{x}} | \boldsymbol{z})$.

A VAE is a generative model, due to the inclusion of a prior distribution $p(z)=\mathcal{N}(0,1)$, which can be sampled to generate synthetic data. Since we approximate $p(z)$ with a unit Gaussian, we then choose to model the encoding process as a multivariate Gaussian with diagonal covariance, $E_{\phi}(\boldsymbol{x}) = \mathcal{N}(\mu_z, \sigma_z^2)$. To enable training of this network we make use of the reparameterisation trick, which allows for the back propagation of gradients in stochastic gradient descent. To do this, instead of $E_{\phi}(\boldsymbol{x})$ directly encoding $\boldsymbol{z}$, we instead encode the means and covariances of each input, $\mu_z$ and $\sigma_z^2$ (in practice, we encode $\ln{(\sigma_z^2)}$ as it is more stable numerically). We then re-sample $\boldsymbol{z}$ in the following way:

\begin{equation}
\label{equ:repara_trick}
    \boldsymbol{z} = \mu_z + \sigma_z \circ \mathcal{N}(0,1)
\end{equation}

\noindent where $\circ$ is an element-wise multiplication and $\mathcal{N}(0,1)$ is a unit Gaussian. The objective function of this VAE is called the Evidence Lower Bound (ELBO), and we train the network by maximising this function. The ELBO has a general form of:

\begin{equation}
    \log p_{\theta}(\boldsymbol{x}) \geq \mathbb{E}_{q_{\phi}(\boldsymbol{z} | \boldsymbol{x})} \log p_{\theta}(\hat{\boldsymbol{x}} | z)-\operatorname{KL}\left(q_{\phi}(z | \boldsymbol{x}) \| p(z)\right)
\end{equation}

\noindent The first term on the right hand side is interpreted as the reconstruction loss by transforming a latent representation $\boldsymbol{z}$ to $\boldsymbol{\hat{x}}$, such as the mean squared error (MSE) or binary cross entropy (BXE) multiplied by the dimensionality of the input data (D). The second term is the Kullback-Leiber (KL) divergence between the encoded representation of $\boldsymbol{z}$ and the prior distribution. In practical terms, the first term optimises the network to produce good reconstructions of the inputs, while the second term acts as a regulariser that punishes the network if the distribution of $\boldsymbol{z}$ drifts from a unit Gaussian.

We can rewrite the ELBO function into an exact loss function, $L_{\text{Total}}$, that our network can optimise in the following way:

\begin{equation}
    L_{\mathrm{Total}} = L_{\mathrm{Recon}} + L_{\mathrm{KL}}
\end{equation}
\noindent where
\begin{equation}
    L_{\mathrm{Recon}}=\mathbb{E}_{q_{\phi}(z | \boldsymbol{x})} \log p_{\theta}(\hat{\boldsymbol{x}} | z) = \left[\mathrm{MSE\ or\ BXE} \right] \times D
\end{equation}
\noindent and
\begin{equation}
    L_{\mathrm{KL}} = -\frac{1}{2}  \sum\limits_{j}\left( 1 + (\log{\sigma_j)^2} - \mu_j^2 - \sigma_j^2 \right)
\end{equation}

When the network is trained, we can use it in a generative way by simply taking a random sample from $p(z)=\mathcal{N}(0,1)$, and feeding it as an input to the decoder network. Alternatively, we can generate latent representations of our inputs and perform a clustering analysis, similar to a conventional autoencoder.

One major problem with VAEs in general is that the regularising effect of the KL term tends to negatively impact the reconstruction quality. With no adjustments, this typically results in blurrier reconstructions than a conventional autoencoder with the same architecture. There is a balance in how well the VAE can learn and disentangle the latent dimensions and how well it can use that space to reconstruct inputs. A simple way to address this balance is by introducing a weight on the KL term, either increasing its effect to improve disentanglement of image properties, or decreasing its effect to improve image quality at the cost of poorer sampling. So called $\beta$-VAE networks, and their successors, have been studied widely, and we will discuss later how we include this in our architecture.

\subsection{VaDE and s3VDC}
\label{sec:s3vdc}

The example VAE given in \cite{2013arXiv1312.6114K} uses only a single unit Gaussian component as its prior on $\boldsymbol{z}$, but the framework can be generalised to include a mixture of Gaussians. While there have been many examples of this, the two most prominent are VaDE (J16) and GMVAE \citep[Gaussian Mixture Variational Autoencoders,][] {2016arXiv161102648D}, which vary in their implementations. The essential difference is that VaDE calculates a single encoded mean and variance for each sample, and learns the GMM means, variances and weights as trainable parameters, whereas GMVAE approximates the GMM means and variances with additional neural networks and keeps the component weights fixed. Recently, C20 presented a study of the state-of-the-art variational deep clustering methods and highlighted some key problems with the frameworks in terms of simplicity, scalability and stability.

For AstroVaDEr, we choose to implement the VaDE algorithm with the optimisations from s3VDC. We will now discuss the modifications to the ELBO and loss functions to turn VAE into VaDE, and the steps taken to incorporate the s3VDC framework. To begin, let us describe the clustering model which is parameterised by the categorical distribution $p(c)$:

\begin{equation}
    c \sim \operatorname{Cat}(\boldsymbol{\pi}), \quad \text {s.t.:} \sum_{c=1}^{C} \pi_{c}=1,
\end{equation}

\noindent where $\pi_c$ are the weights of the clusters. We find that in practice \smash{$\sum_{c=1}^{C} \pi_{c}=1$} is not always true when the model is training, but experiments invetigating constraining these values via normalisation or a softmax function did not produce improved results. With this categorical distribution, we also change how the latent variable $\boldsymbol{z}$ behaves with the following prior:

\begin{equation}
    \boldsymbol{z} \sim \mathcal{N}(\mu_c, \sigma_c^2\boldsymbol{I}),
\end{equation}

\noindent where $\mu_c$ and $\sigma_c^2$ are the means and diagonal covariances of the clusters, and $\boldsymbol{I}$ is the identity matrix. Under VaDE, $\pi_c$, $\mu_c$ and $\sigma_c$ are all trainable weights within a neural network. The ELBO objective of VaDE is then given as:

\begin{equation}
    L_\text{Total} = \mathbb{E}_{q_{\phi}(\boldsymbol{z} , c|\boldsymbol{x})} \log p_{\theta}(\hat{\boldsymbol{x}} , z)-\operatorname{KL}\left(q_{\phi}(z , c|\boldsymbol{x}) \| p(z,c)\right)
\end{equation}

\noindent As before, the first term in equation 10 is the reconstruction term, now given with respect to $q_{\phi}(\boldsymbol{z},c|\boldsymbol{x})$, instead of $q_{\phi}(\boldsymbol{z} | \boldsymbol{x})$. The KL term now describes the Kullback-Lieber divergence between the latent variable $\boldsymbol{z}$ and the cluster model $p(z,c)$. Within the neural network model, the KL component of the loss is calculated as follows:

\begin{multline}
    L_\text{VaDE}(x_i) = -\frac{1}{2} \sum\limits_{j=1}^{J} \left(1 + \log {{\sigma_{zj}^i}^2}\right) \\
    - \sum\limits_{c=1}^{K} \left( \gamma_c^i \log (\pi_c) + \gamma_c^i \log (\gamma_c^i)\right) \\
    + \frac{1}{2} \sum\limits_{c=1}^{K} \left( \gamma_c^i \sum\limits_{j=1}^{J} \left(\log (\sigma_{cj}^2) + (\mu_{zj}^i - \mu_{cj})^2 + \left(\frac{{\sigma_{zj}^i}}{\sigma_{cj}}\right)^2\right) \right) \\
    + J\log 2\pi,
\end{multline}

\noindent where $J$ is the dimensionality of the embedded spaces, $K$ is the number of cluster components, $x_i$ is an input sample, $\mu_z^i$ and ${\sigma_z^i}^2$ are the embedded mean and covariance representations of $x_i$ from the encoder network, and $\gamma_c^i$ is the cluster probability of the input $x_i$. To calculate the cluster probabilities of an input sample, we follow suit with C20 and use the {\tt scikit-learn} {\it Python} implementation \citep{scikit-learn}.\footnote{The full calculation of $\gamma_c^i$ can be found in the Gaussian Mixture source code for \href{https://scikit-learn.org/stable/modules/generated/sklearn.mixture.GaussianMixture.html}{\tt scikit-learn} and we also refer the reader to the github repository for C20: \url{https://github.com/king/s3vdc}.}

\begin{equation}
    \log(p(c,z)) = - \log(p(z,c)) - \log(p(c))
\end{equation}
\begin{equation}
    \gamma_c^i = \frac{e^{\log(p(c,z))}}{\sum\limits_{c=1}^{K}e^{\log(p(c,z))}}
\end{equation}

Training of VaDE then follows the usual scheme for a VAE. A batch of samples $\boldsymbol{x}$ is embedded into the latent space represented by $\boldsymbol{\mu}_z$ and $\boldsymbol{\sigma}_z^2$. We use the reparameterisation trick to sample $\boldsymbol{z}$, which is fed into a probabilistic decoder. The main difference is that the network is now conditioning the latent space such that $\boldsymbol{z}$ tends towards $\mathcal{N}(\mu_c, \sigma_c^2\boldsymbol{I})$ instead of $\mathcal{N}(0,1)$.

C20 introduced a number of improvements to variational deep clustering methods which we implement in AstroVaDEr. First, instead of pretraining the network without a KL component, as done in VaDE, we use an $\alpha$-training\footnote{C20 call this a $\gamma$-training phase, but we wish to avoid confusion with the cluster probability, $\gamma_c$.} phase which incorporates a low weighted KL regulariser from a simple $\mathcal{N}(0,1)$ prior for $T_\alpha$ epochs. This pretraining primes the network to produce good reconstructions, without wandering too far away from the latent prior such that the embedded space becomes too unstructured for good clustering fits. During this phase we weight the KL loss component by a factor $\alpha = 0.0005$, such that the network loss at epoch $t$ is calculated as:

\begin{multline}
    L_{\mathrm{Total}} = L_{\mathrm{Recon}} + \alpha L_{\mathrm{KL}}, \\
    T_\alpha \geq t > 0
    \label{equ:KL_Loss}
\end{multline}

The next optimisation is a phased annealing program. The GMM prior is introduced, but the contribution to the total loss is slowly increased during the `annealing phase' using a weighting factor $\beta$. The ramping up of $\beta$ follows a polynomial function until $\beta = 1 + \alpha$. Following annealing we  train the model in a `static phase' with equal weighting given to the reconstruction and KL divergence losses. These two phases are repeated a fixed number of times. The reason for this is to hamper the effects of the competing losses, wherein the reconstruction loss is pressuring the model to use as much of the latent space as possible to improve image recovery, while the KL loss is trying to force the embedded representations onto the prior distribution. By phasing the weight of the KL loss, we can improve the disentanglement between the latent variables, without significantly reducing reconstruction quality. The annealing and static phases are repeated for $M$ periods, lasting $T_\beta+T_s$ epochs each, where $T_\beta$  and $T_s$ are the number of epochs in the annealing and static phases respectively. The network losses during these phases are:

\begin{multline}
    L_{\mathrm{beta}} = (\beta^u + \alpha) L_{\mathrm{KL}} \\
    \text{s.t.:} \beta = \left[  \frac{t - T_\alpha - (m-1)(T_\beta+T_s)}{T_\beta} \right], \\
    T_\alpha + (m-1)(T_\beta + T_s) + T_\beta \geq t > T_\alpha + (m-1)(T_\beta + T_s)
    \label{equ:vade_Loss}
\end{multline}
\noindent and
\begin{equation}
    L_{\mathrm{static}} = L_{\mathrm{Recon}} + L_{\mathrm{KL}},
\end{equation}

\noindent where $t$ is the current epoch, and $u$ is a polynomial factor that dictates the rate that $\beta$ increases to $1$.

The third recommendation from C20 concerns the initialization of the GMM weights. Naively, one might assume to simply initialise the weights randomly, but this would require spinning up numerous versions of the model to find the best fit, which is time consuming and requires a large amount of computing power. A common technique is to instead calculate the initial weights by fitting a GMM onto the embedded representations of the training data after some fixed pretraining period. C20 opt for this approach, but they do so by investigating the number of input samples required to perform a satisfactory fit. In what they call `mini-batch GMM Initialization,' we simply fit a GMM using $k$ batches of size $L$, instead of the full data set.

Finally, C20 address the problem of {\tt NaN} losses in VDC models. The most common culprit of {\tt NaN} values in these networks can be traced back to the approximation of $\log p(c,z)$ and $\gamma_c$. Essentially, it is possible that a sample's probability within a cluster falls so close to zero that it generates a {\tt NaN} or {\tt Inf} value within the loss function. C20 address this by introducing a min-max scaling of $\log p(c,z)$, which prevents its values from getting too small. Considering $\boldsymbol{V}$ as the full matrix representation of $\log p(c,z)$, we calculate a scaled value $\hat{\boldsymbol{V}}$ with a range $[-\lambda, 0]$:

\begin{equation}
    \hat{\boldsymbol{V}} = \lambda \frac{\left[ \boldsymbol{V} - \operatorname{min} \boldsymbol{V} \right]}{\left[ \operatorname{max} \boldsymbol{V} - \operatorname{min} \boldsymbol{V} \right]}
    \label{eqn:minmax}
\end{equation}

With these optimizations in mind, we will now turn our attention to the neural network architecture which we use to construct the encoders and decoders that will be used in our experiments.

\subsection{AstroVaDEr Architechture}

\begin{figure*}
 \includegraphics[width=\textwidth]{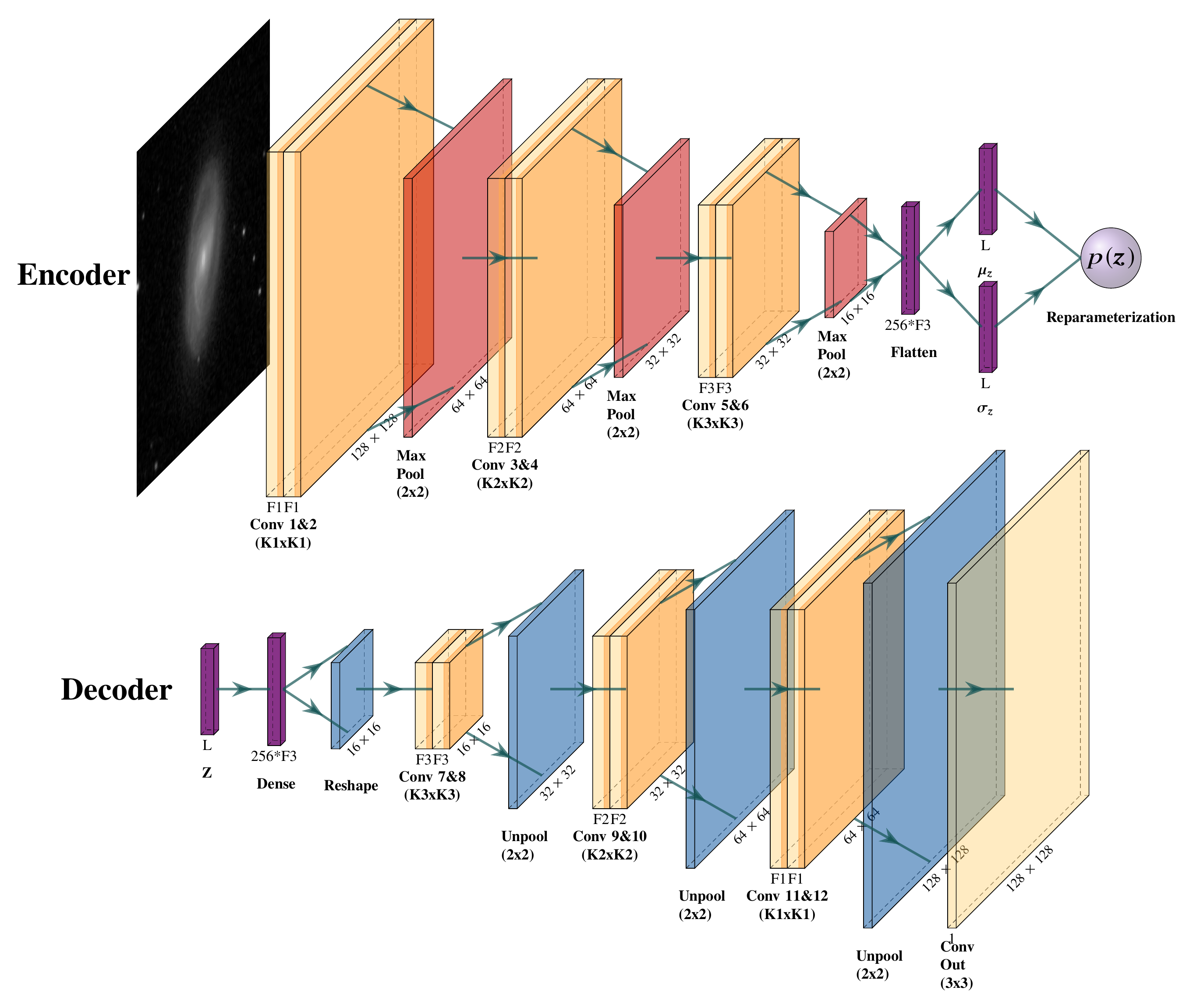}
 \caption{Diagrams representing the network architecture of AstroVaDEr. Shown on the top row is the Encoder network, $E_{\phi}(\boldsymbol{x}) = \boldsymbol{z}$, and the bottom row is the Decoder network, $D_{\theta}(\boldsymbol{z}) = \boldsymbol{\hat{x}}$. The core layers of $E_{\phi}(\boldsymbol{x}) = \boldsymbol{z}$ and $D_{\theta}(\boldsymbol{z}) = \boldsymbol{\hat{x}}$ consist of paired convolutional layers, each with ReLU activation, we use max pooling layers in the Encoder and Upsampling layers in the Decoder to decrease and increase the dimensionality, respectively. The paired convolutional layers have $(64,64,16)$ filters, with kernel sizes of $(3,5,5)$. We utilise layer flattening and reshaping to compress and decompress the activation maps in and out of the embedded space. The $L$-unit dense layers represent $\mu_z$ and $log(\sigma_z^2)$, the latent embedding. $p(z)$ represents the reparameterisation trick. We implement a \texttt{Keras} layer to contain the GMM weights at the end of the Encoder.}
 \label{fig:astrovader_architecture}
\end{figure*}

AstroVaDEr is constructed using the \texttt{Keras} Python API \citep{chollet2015keras}. The core architecture is based on the CNN developed by W20 for GZ2. W20 is in turn based on VGGNet \citep{2014arXiv1409.1556S}, and we follow suit with the core of both the encoder and decoder networks following this architecture. Figure \ref{fig:astrovader_architecture} shows the model architecture employed in this work. The main model architecture is designed to be flexible, and the model parameters shown here are simply those that were chosen during our optimisation process, discussed in Section \ref{sec:training}. In practice, the input size, number of convolutional filters, kernel sizes and latent variables can all be tuned for different applications.

The encoder takes an input image, a $128\times128$ pixel gray scale image in our case, and successively reduces the size and learning of the key features with convolutional and pooling layers. A `block' in the encoder has two convolutional layers with a stride of 1 and `same' padding, followed by a $2\times2$ max pooling layer. The encoder has three such blocks, which reduce the image size by a factor of 8 in total. The number of filters in each block, and the kernel sizes, are set during hyperparameter optimisation. The output from the third block is flattened to a one dimensional array. The flattened activation maps are passed into a pair of dense layers, which encode $\mu_z$ and $\log(\sigma_z^2)$. We implement three custom layers, the first performs the reparameterisation trick to sample $z$ from Equation \ref{equ:repara_trick}, followed by a pass-through layer that initialises the GMM weights for training, and finally a layer which calculates the cluster probabilities for each input sample.

The decoder essentially performs the opposite transformation as the encoder. Each block in the decoder consists of two convolutional layers and a bi-linear $2\times2$ up-sampling layer. The embedded code for a sample is fed into a dense layer with the same number of units as the flattened encoder layer, and is then reshaped into an image with bands equal to the number of filters in the last encoder block. The three decoder blocks use the same number of filters as the encoder but in reverse, and the final up-sampled output is fed into a final convolutional layer with either a single filter for gray scale input or three filters for RGB images.

Throughout the network we choose to implement the Leaky ReLU activation function \citep{2019arXiv190306733L}, with a slope of $\alpha=0.1$. We use this function, as opposed to the popular ReLU activation, to tackle the "dying ReLU" problem. Dying ReLU occurs in ReLU activated neurons where it is possible for a neuron to be set to zero and then can no longer contribute to learning. Essentially the neuron `dies' \citep{2019arXiv190306733L}. Leaky ReLU addresses this by allowing a small slope for negative values, thus preventing the gradient going to zero when the activated output is forced to be positive. This activation is used for all convolutional layers in the encoder, the dense layer following the decoder input, and all but the final convolutional layer in the decoder. The embedded dense layers have a linear activation, and the output has a ReLU activation (the output is not affected by the dying ReLU problem because it is the first layer in the back propagation). All convolutional and dense layers implement a l2 regulariser on their weights and biases with a factor of $0.01$.

The GMM weights are fairly well behaved, but we do note that it is useful to have a constraint on the GMM covariances to keep them positive definite. In circumstances where there are too many mixture components, or poorly defined components from the initialisation, we find that the model will attempt to remove those components by shrinking their covariances to zero or even negative values, which results in {\tt NaN} values propagating through the network when an inverse or logarithmic covariance is calculated.

Training is performed on two NVIDIA Tesla V100 graphics processor units (GPUs) on the University of Hertfordshire High Performance Computing cluster, using the \texttt{Keras multi-gpu-model} functionality . We experimented with a variety of batch sizes, and settled on 180 samples per batch split across the two GPUs. The main effects of different batch sizes is in the CPU bottleneck in loading and preprocessing the inputs. We choose to utilise the \texttt{Keras} Image Data Generator class to load each image from the hard disk, perform random transformations and feed batches to the network. If the batch size is too large, the network can struggle to generate batches fast enough for the GPU to process them through the model.

\section{Model Training}
\label{sec:training}

\subsection{Hyperparameter Optimisation}

As stated above, the core architecture of AstroVaDEr is intended to be flexible to different applications. We will now discuss the specific settings which were used to produce the results in this paper. We used Bayesian Hyperparameter Optimisation \cite[via Hyperopt]{hyperopt} \footnote{Specifically, the {\tt hyperas} implementation \url{https://github.com/maxpumperla/hyperas}} to narrow-down the parameter search space and then a manual search for settings that optimised reconstruction and clustering quality.

The network is optimised using the Adam optimiser \citep{2014arXiv1412.6980K} with an initial learning rate of $3\times10^{-4}$, we decrease the learning rate every five epochs using exponential decay, such that the final learning rate is $10^{-6}$. We set the number of latent variables to be $20$. Larger values do improve the reconstruction quality of the network, but for the purpose of this paper the lower number makes interpretation more straightforward. For the convolutional blocks, the encoder uses $64$ filters in the first and second pairs of convolutions, and 16 filters in the final pair. The convolutions use $(3\times3)$ kernels in the first block, and $(5\times5)$ kernels in the second and third blocks. The decoder uses the same configuration, but in reverse. The flattened dimensionality of the encoder before the latent embedding is thus $16\times16\times16=4096$, which also corresponds to the number of units in the first fully-connected layer in the decoder.

The s3VDC framework introduces an additional set of hyperparameters that must be optimised. These include: number of GMM mini-batches, $\alpha$-training epochs, annealing phase epochs, static training epochs and annealing-static periods. These parameters pose a challenge for Bayesian optimisation however, as they require full end-to-end runs of model training, each of which can take upwards of 12 hours to complete. Instead, we used a purely qualitative approach to setting these parameters, which first involved varying the length of the $\alpha$-training period until we were satisfied with the quality of the reconstructions, and then again manually adjusting the length and number of annealing periods until we found convergences in the GMM weights. We show the results of this search in Table \ref{s3VDCOpt}. Future work will attempt to quantify and streamline this process, by investigating different early stopping conditions at different stages of training and analysing how the number of epochs in each phase is related to the learning rate and the weighting factors on the KL divergence. One additional outcome of this search was that $L_{\mathrm{KL}}$ often overpowered the reconstruction losses under certain combinations of the number of Gaussian components and latent variables, even when we employed the annealing periods. This feature of the training requires further study to fully understand, but for the purposes of this work we found that including an additional weighting of $0.3$ on $L_{\mathrm{KL}}$ allowed for good reconstruction quality and clustering results.

\begin{table}
    \centering
    \caption{s3VDC hyperparameters found by manual search.}
    \label{s3VDCOpt}
    \begin{tabular}{ll}
        \textbf{Hyperparameter} & \textbf{Value} \\ \hline
        $\alpha$-training epochs & 100 \\
        GMM Batches & 200 \\
        Annealing epochs & 25 \\
        Static epochs & 25 \\
        \# Annealing Periods & 5
    \end{tabular}
\end{table}

\begin{figure*}
    \centering
    \includegraphics[width=0.95\textwidth]{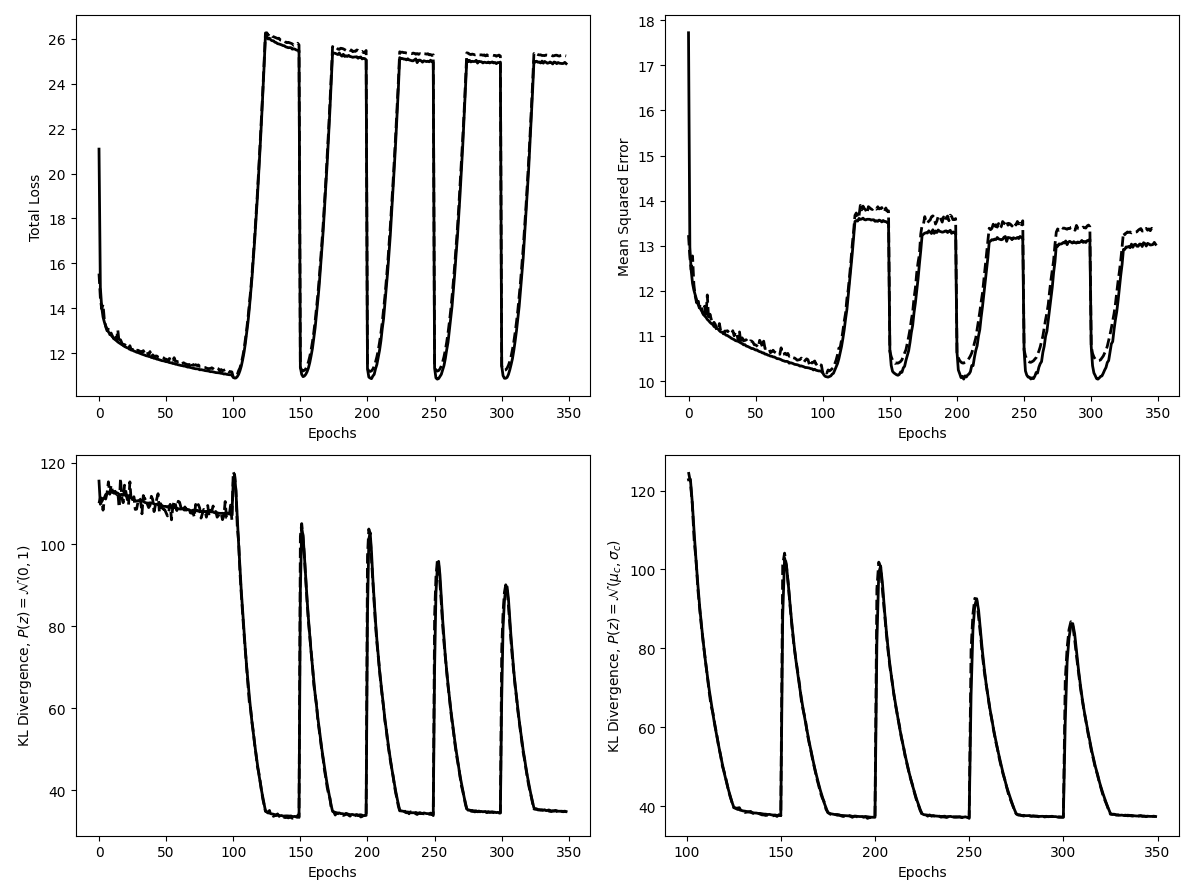}
    \caption{Training (solid lines) and Validation (dashed lines) losses for AstroVaDEr obtained while training the network using the hyperparameters discussed in this Section. Top left is the total loss throughout training, calculated from Equations \ref{equ:KL_Loss} \& \ref{equ:vade_Loss} before and after 100 epochs, respectively. This loss includes additional penalties to training from regularisation of the weights and biases. Top right shows the reconstruction loss, measured as the total mean squared error per sample. Bottom left shows the KL divergence associated with the vanilla VAE prior, $p(z)=\mathcal{N}(0,1)$, which is used in the first $100$ epochs. Bottom right shows the KL divergences from the GMM prior, $\boldsymbol{z} \sim \mathcal{N}(\mu_c, \sigma_c^2\boldsymbol{I})$, from epoch $101$ onward, which is annealed following the procedure outlined in Section \ref{sec:s3vdc}.}
    \label{fig:Training Losses}
\end{figure*}

The final parameter we can control is perhaps the most important given our final goal of producing a generative classifier: the number of components to include in the Gaussian mixture. The number of clusters to use in any unsupervised clustering task depends on the data set, the method and the desired result. For some problems one may know in advance the number of clusters the data naturally falls into, for example in clustering the MNIST \citep{lecun-98} handwritten digits data set one knows there should be ten clusters. However, as is the case with real world applications of these algorithms, we often do not have a priori knowledge of the ideal number of clusters for describing the data. This is further complicated by our goal of developing a generative model, which will be able to selectively generate synthetic images from a particular component in our Gaussian mixture. We have to balance, in essence, the  philosophical question of `how many types of galaxy are there?' with the practical question of `how many components can the network accurately model?' Too few clusters and we risk blurring the definitions of different morphological types and losing the ability to distinguish between similar classes. With too many clusters the network will begin to encounter sparsity problems where poorly populated components are effectively turned off.

Comparing to other astronomical works, we see a wide variety in approaches to this problem. For example, approaching the problem from the perspective of supervised learning we may simply attempt to define two clusters, following a binary classifier such as in W20. The problem here is that with a CNN classifier one can define the specific classes to be determined (e.g.\ spiral or elliptical, barred or unbarred), but we do not have this luxury. Two clusters may instead define edge-on versus face-on galaxies, or isolated galaxies versus crowded images. At the other end of the scale is the work of \cite{2020MNRAS.491.1408M}, who use as many as 160 clusters in their unsupervised algorithm. There are a few key differences here, firstly \cite{2020MNRAS.491.1408M} use a Growing Neural Gas model and hierarchical clustering to produce learned feature vectors for each object (which are analogous but not equivalent to our latent space) and each object is assigned a cluster based on {\it k}-means clustering which typically does not fail to produce large numbers of roughly equally populated clusters. A GMM, especially as implemented in VaDE, is prone to model collapse due to exploding and vanishing covariances and vanishing weights on the individual components.

Two recent works in astronomy use similar techniques to ours to solve different problems: \cite{2020MNRAS.494.3750C} use a CAE trained on simulated images of strong gravitational lenses, followed by a Bayesian GMM applied to the embedded samples post training to produce a classifier. They use a number of clusters in their mixture model equal to the number of latent dimensions, which in a simplified way can be explained as saying that each variable in the latent space (i.e. each unit in the fully connected layer) approximates to one feature and therefore one cluster. The assumption that each variable only controls one feature, however, is not always true, which we explore in Section \ref{sec:generative}. \cite{2019PASP..131j8011R} instead apply a CAE to images of radio galaxy images in order to classify their morphologies, and produce clusters based on a Self Organising Map (SOM) and {\it k}-means clustering of the latent space. They show clustering results for 4 and 8 clusters, but stress that those numbers were chosen as a general demonstration of the $20\times20$ SOM for clustering purposes. It should be emphasised that both of these methods cluster the data on the latent space after training on conventional autoencoder architectures, while we are attempting to perform the task during training on a generative architecture.

We tested a range of techniques to find optimal cluster numbers, which we will briefly explain before discussing our chosen method. C20 demonstrate how they optimise the number of clusters for a dataset where the true number of clusters in unknown. They train their network fully with a range of values for the number of components in the GMM, and then compare various unsupervised clustering metrics. They use the Calinski-Harabasz \citep[CH,][]{doi:10.1080/03610927408827101} index, a measure of the ratio between in inter-cluster dispersion and intra-cluster dispersion that is higher when clusters are dense and well separated, and the simplified silhouette score \citep{ROUSSEEUW198753}, which measures how similar samples are to their cluster members compared to samples outside their cluster and ranges between $-1$ and $1$ where higher numbers suggest better clustering. They also investigate the disentanglement of the clusters by projecting the embedded data into two dimensions with t-distributed stochastic neighbour embedding \citep[TSNE,][]{10.5555/2968618.2968725} and the marginal likelihood of the GMM over the dataset. C20 suggest that comparing all of these factors together can determine the best number of clusters to use. We explored this option, but the main difficulty with these methods when working with very complex data is that training just an individual instance of the network takes many hours. Training the network multiple times with different clustering settings end-to-end isn't computationally efficient \citep[especially when we consider the already substantial environmental impact of high performance computing,][]{2020NatAs...4..819P}. Added to that, the experiments we did perform were largely inconclusive due to the high overlap between the learned GMM components (see Section \ref{sec:Clustering} for details).

However, we settled on what seems a fairly intuitive methodology that finds an optimum number of clusters without needing to train multiple versions of the full AstroVaDEr network, which exploits the behaviour of the {\tt scikit-learn} \citep{scikit-learn} Bayesian Gaussian Mixture (BGM) method. BGM is a variational GMM and uses the expectation-maximisation technique \citep{10.2307/2984875} to maximise the ELBO of the log-likelihood \citep{kullback1951, Attias2000, McLachlan1997, Bishop2006}. This method can be employed in a mode which approximates an infinite-mixture model with a Dirichlet Process \citep{ferguson1973}. In this mode, the algorithm naturally sets low probability clusters to have zero contribution to the mixture, essentially setting the number of components automatically. How this is achieved can be understood with the stick-breaking analogy; consider a unit-length stick that represents the Dirichlet Process, and step by step we break off a piece of that stick. Each broken piece of stick represents a groups of points that fall within the mixture, and has a random length that defines the weight of the piece. The remaining length of stick is successively broken down, and the last piece is used to represent the the points that don't fall into the other pieces. In this way, as long as we choose a number of components that is greater than the true optimum number of clusters, the excess components will have their weights set to zero and are removed from the model. We used this behaviour to choose the number of clusters to fit to our data by initialising many instances of a BGM with different mini-batches of embedded samples, and found that $12$ clusters emerged as a fairly consistent mixture configuration. Compared to the other methods discussed here, using this stick-breaking method is computationally fast, consistent, intuitive, and crucially doesn't require comparing clustering metrics and visualisations of a dozen fitted models where the best options aren't immediately apparent.

\subsection{Training Losses}

For this work, we trained AstroVaDEr with the hyperparameters discussed previously and show the losses obtained during the training in Figure \ref{fig:Training Losses}. Losses from the training data and validation data are shown with solid and dashed lines, respectively, and in all cases are the mean values over all batches. For the total, reconstruction and vanilla VAE KL divergence we show the values across all $350$ epochs, and for the GMM KL divergence we show its values from epochs $101-350$ as we do not train the GMM parameters in the first $100$ epochs. The $\alpha$, annealing and static training phases are all recognisable in the plots, with the $\alpha$ training taking the first $100$ epochs and then the other phases alternating as previously described.

In the top left of Figure \ref{fig:Training Losses} we show the total loss calculated from Equations \ref{equ:KL_Loss} \& \ref{equ:vade_Loss} plus the penalties associated with the L2 regularisers on the model weights and biases. In the first $100$ epochs the total loss approaches, but does not fully reach, a minimum convergence, this suggests that we could potentially train in this phase longer, but that runs the risk of collapsing the latent space onto the unit Gaussian prior too much. As the annealing factor on the GMM loss increases during the $\beta$-annealing phase, the total loss increases with the cubic factor on the weight, and then decreases slowly in the first static phases before plateauing in the later ones. By epoch $350$, the total loss has flattened in the final static phase, and the minimum total loss achieved when the annealing factor is at its minimum does not appreciably change through the training.

The reconstruction losses, measured as the mean squared error per sample in the training and validation sets, are shown in the top right plot of Figure \ref{fig:Training Losses}. The error here follows a similar pattern to the total loss, decreasing steadily during the $\alpha$-training phase and then increasing-plateauing-minimising as the network cycles through the $\beta$ and static training phases. Unlike the total losses, we do see some reduction in the peak reconstruction error with each annealing phases, and additional training may result in better imaging quality at the expense of increased computational time. The exact effect of the static phases on the reconstruction quality is not overly clear from our study, as it does not seem to change during those epochs. According to C20, this is a fine-tuning stage where the network should be attempting to find an equilibrium between the clustering and reconstructions, but further testing is needed to be more specific.

The last two panels show the KL divergence for the unit Gaussian and GM priors on the left and right, respectively. By design due to the small weigh applied during $\alpha$-training, the unit Gaussian divergence does not fall a large amount. During the annealing periods the unit Gaussian loss is not used in the training, but we track it to compare with the GMM loss. As the weighting factor on the GMM loss increases, the loss decreases to a minimum where it remains steady during the static training. We do note, that like the peak reconstruction loss decreasing with each annealing period, so too does the peak GMM divergence when the static phases end and the weighting factor is reset close to zero.

\section{Results}
\label{sec.results}

\subsection{Image Reconstruction Quality}

The first step in assessing the quality of the trained model is to examine its ability to reconstruct the test images that were not included in the training sample. We draw from the test dataset, defined in Section \ref{sec.Data}, for this purpose. We assess the reconstructions in two ways, first by a qualitative visual inspection, and second by investigating the objects that the network performs best and worst at reproducing. Reconstructions are produced by passing test images, without any augmentations, into the AstroVaDEr network, where they are embedded into the latent space by the encoder and then retrieved by the decoder.

\begin{figure*}
    \centering
    \includegraphics[width=0.90\textwidth]{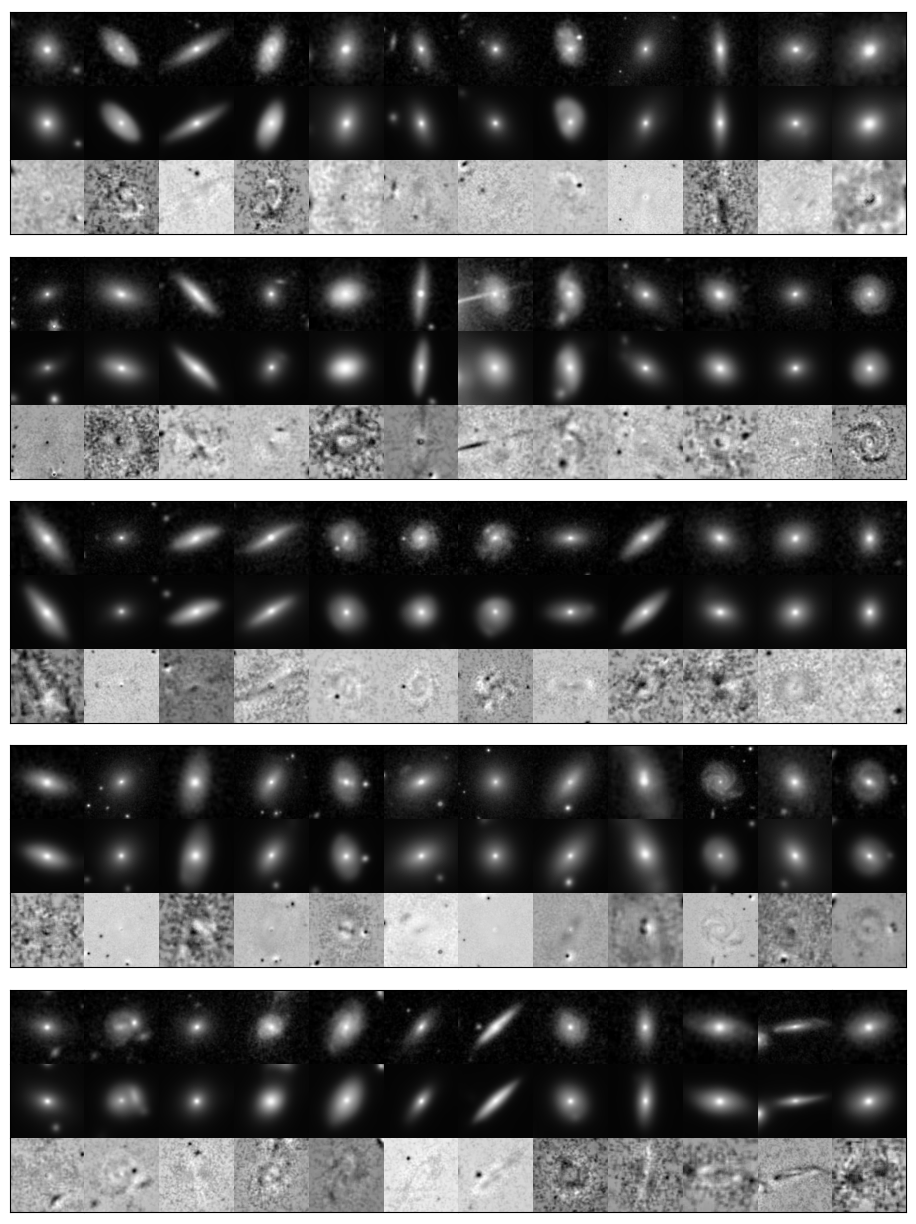}
    \caption{Image reconstructions of a random sample of galaxies. Each galaxy is shown with the non-augmented, gray scale input image, followed by the AstroVaDEr reconstruction and then the residual between the input and reconstruction. Input images and reconstructions are shown with a linear scale and pixel range of $(0,1)$.}
    \label{fig:image_recons}
\end{figure*}

\begin{figure*}
    \centering
    \includegraphics[width=\textwidth]{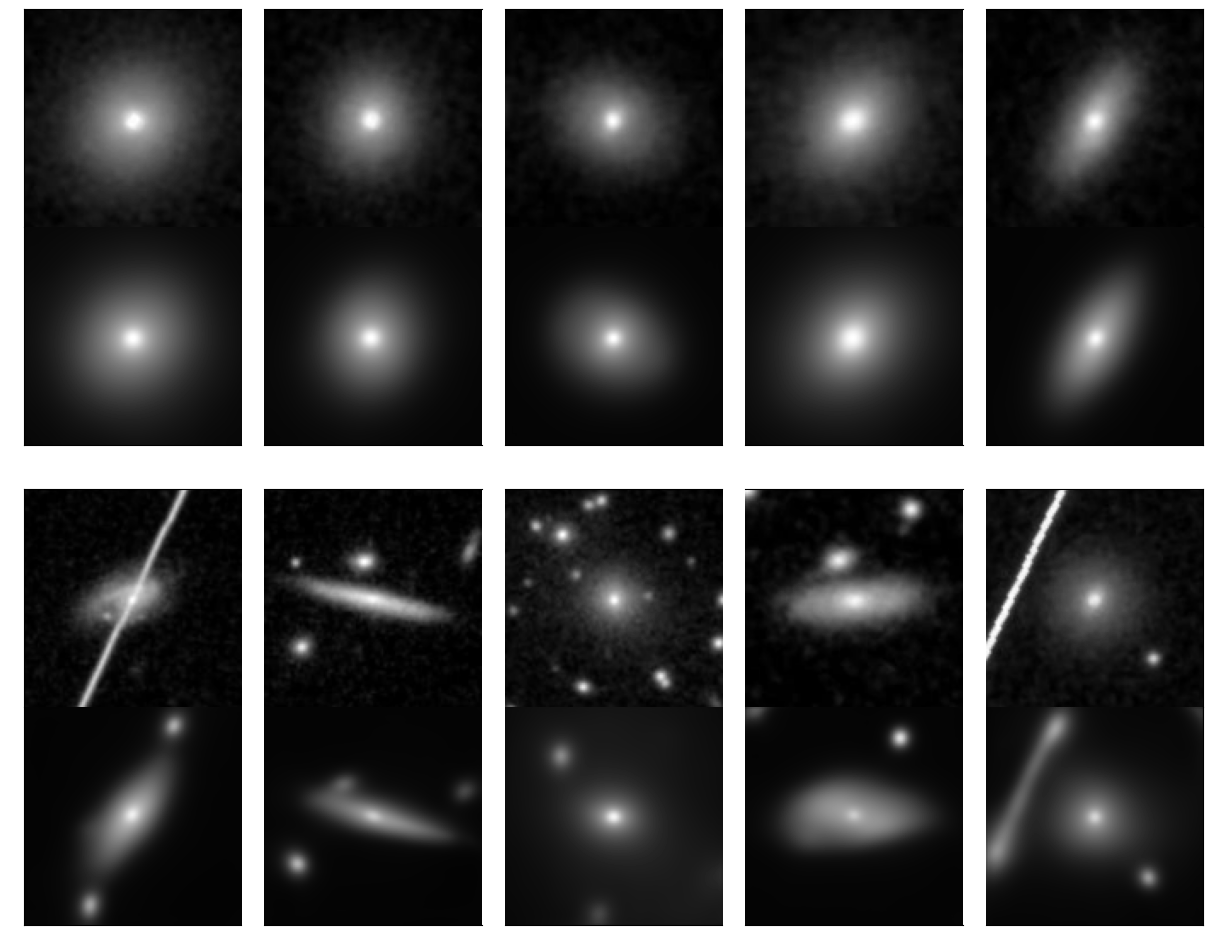}
    \caption{Comparison of reconstruction quality between galaxies with low reconstruction loss (top two rows), and high reconstruction loss (bottom two rows). The galaxies were selected by calculating the mean squared error between the input images and reconstructions, and choosing those with the five galaxies with the lowest error and highest error.}
    \label{fig:recons_mse}
\end{figure*}

Figure \ref{fig:image_recons} shows the random selection of galaxies, with their input images, reconstructions, and residuals. The residuals provide a better idea of what parts of the galaxies the network is performing well at reconstructing and where it struggles. Broadly speaking, the overall shape of each galaxy is well-preserved by the network, by which we mean that global properties such as axis ratio/inclination angle, surface brightness, angular size, and orientation, are all visually reproduced. VAEs like AstroVaDEr have a the well known property of producing blurry or `fuzzy' images, which we see here as well. We can see in the reconstructions and residuals that the network is effectively denoising the input image, however the finer details are smoothed out. The result is that the network struggles to reproduce fine-scale morphological features like spiral arms and bars. Inspecting some of the highly featured objects in Figure \ref{fig:image_recons}, we can clearly see that the signatures of spiral arms, bars and rings all appear in the residuals.

Addressing the blurriness of VAE generated images (both reconstructions and synthetic images), has been an ongoing area of research, and progress has been made \citep{2017arXiv170603643Y,2018arXiv181207238A,2019arXiv190305789D,2020arXiv200207514A,2020arXiv200512573K}. The key issue, as discussed in \cite{2020arXiv200207514A}, appears to be in the capacity and sparsity in the network. There needs to be sufficient capacity, i.e.\ the number of latent variables, to contain the necessary information for reconstruction, but one must also prevent the collapse of those variables, which induces sparsity in the sense that the network tries to use as few of the latent variables as possible.

Tackling this problem is a matter of carefully balancing the reconstruction loss and the KL divergence between the latent embedding and the prior distribution. Essentially, as the network begins training it will force $\sigma_z$ to be as close to $0$ as possible, ensuring a high degree of confidence in $\boldsymbol{z}$ and therefore $\boldsymbol{\hat{x}}$. However, as the reconstruction loss decreases, and the network focuses more on optimising the KL divergence (in our case, between the embedding and the learned GMM), it instead increases $\sigma_z$ to improve the overall coverage of the latent space. With a less certain measure of $\boldsymbol{z}$, there is a subsequent loss in the reconstruction quality. Solving this problem with AstroVaDEr in its current state is beyond the scope of this work, this is because the techniques that have been developed to do so were developed on `vanilla' VAEs with a unit Gaussian prior. Future work in improving AstroVaDEr will involve integrating one or more of the recent developments in balancing the competing losses with the GMM prior. 
%For now, we shall continue with the network as is.

We have already seen that the network struggles with finer structural details within the reconstructions. To further investigate the strengths and weaknesses of the network, we present Figure \ref{fig:recons_mse}. We calculate the mean squared error for each galaxy in the test set and present the highest and lowest error objects. The top two rows show input and reconstruction images of the five galaxies with the lowest mean squared error. As you may expect from Figure \ref{fig:image_recons}, these images are smooth galaxies; they have a variety of axis ratios and surface brightness profiles, but lack high spatial frequency features like spiral arms. The objects with the highest reconstruction errors are shown in the bottom row, and reveals the two broad categories that AstroVaDEr struggles with the most in terms of mean squared error. The high error objects include those that have visible artifacts and stars, as well as objects that are crowded by multiple nearby sources, be they foreground, background or interacting. One could argue that it would make sense to utilise the Galaxy Zoo labels to remove objects that have imaging artifacts. That would not only defeat the purpose of approaching this task completely unsupervised, but, as we will see in the next section, these objects are {\it identifiable} by the network despite their low reconstruction quality, and this result can be exploited in other ways, such as automatic artifact detection.

\begin{table*}
    \centering
    \begin{tabular}{c|c|c|c|c|c|c|c|c|c|c|c|c|c}
        \textbf{Mixture Component} & 0 & 1 & 2 & 3 & 4 & 5 & 6 & 7 & 8 & 9 & 10 & 11  \\ \hline\hline
        \textbf{No. Objects} & 1727 & 1344 & 6455 & 5771 & 3166 & 2462 & 2781 & 6313 & 4622 & 849 & 3727 & 1943 \\ \hline
        \textbf{No. Objects with $\gamma_c > 0.1$} & 3935 &  2674 & 19154 & 19290 &  5160 &  5902 &  5812 & 16389 & 17214 & 2065 & 19748 &  4180 \\ \hline
        \textbf{No. Objects with $\gamma_c > 0.3$} & 1956 & 1459 & 3406 & 4476 & 3213 & 2516 & 2722 & 4937 & 3243 & 1040 & 2721 & 2155 \\ \hline
        \textbf{No. Objects with $\gamma_c > 0.5$} & 1287 & 1011 &    0 & 2132 & 2459 & 1699 & 1758 & 2331 &  876 &  700 &    0 & 1476 \\ \hline
        \textbf{Mean Silhouette Score} & -0.151 & -0.132 & 0.112 & -0.063 & -0.215 & -0.185 & -0.195 & -0.078 & -0.094 & -0.257 & 0.034 & -0.160 \\ \hline
    \end{tabular}
    \caption{The number of objects from the test set in each mixture component based on the highest likelihood component for each object.}
    \label{tab:cluster_assignmnts}
\end{table*}

The image reconstructions presented here are a limited sample of the 41,000 images in the test dataset. However, this qualitative analysis demonstrates the reconstructive abilities of the network and highlights important areas of improvement. Of course, the reconstructive ability of AstroVaDEr is not the primary goal, and we will now look at the first of the two main tasks of the network: unsupervised clustering.

\subsection{Unsupervised Clustering Results}
\label{sec:Clustering}

Assessing the quality of an unsupervised clustering results is a difficult task, especially given that we have no exact ground truth label to compare our results to. This is perhaps made more difficult by the fact that AstroVaDEr is probabilistic, owing to its variational inference framework and GMM prior. Therefore we do not get a fixed cluster label for each object, but rather a likelihood that an object is drawn from a particular component in the mixture. We can use this to our advantage by inspecting galaxies based on their most likely component, for which we assign a cluster label, and by looking at the most probable objects in each component.

\begin{figure*}
    \centering
    \includegraphics[trim = 10mm 10mm 10mm 10mm, clip, width=1\textwidth]{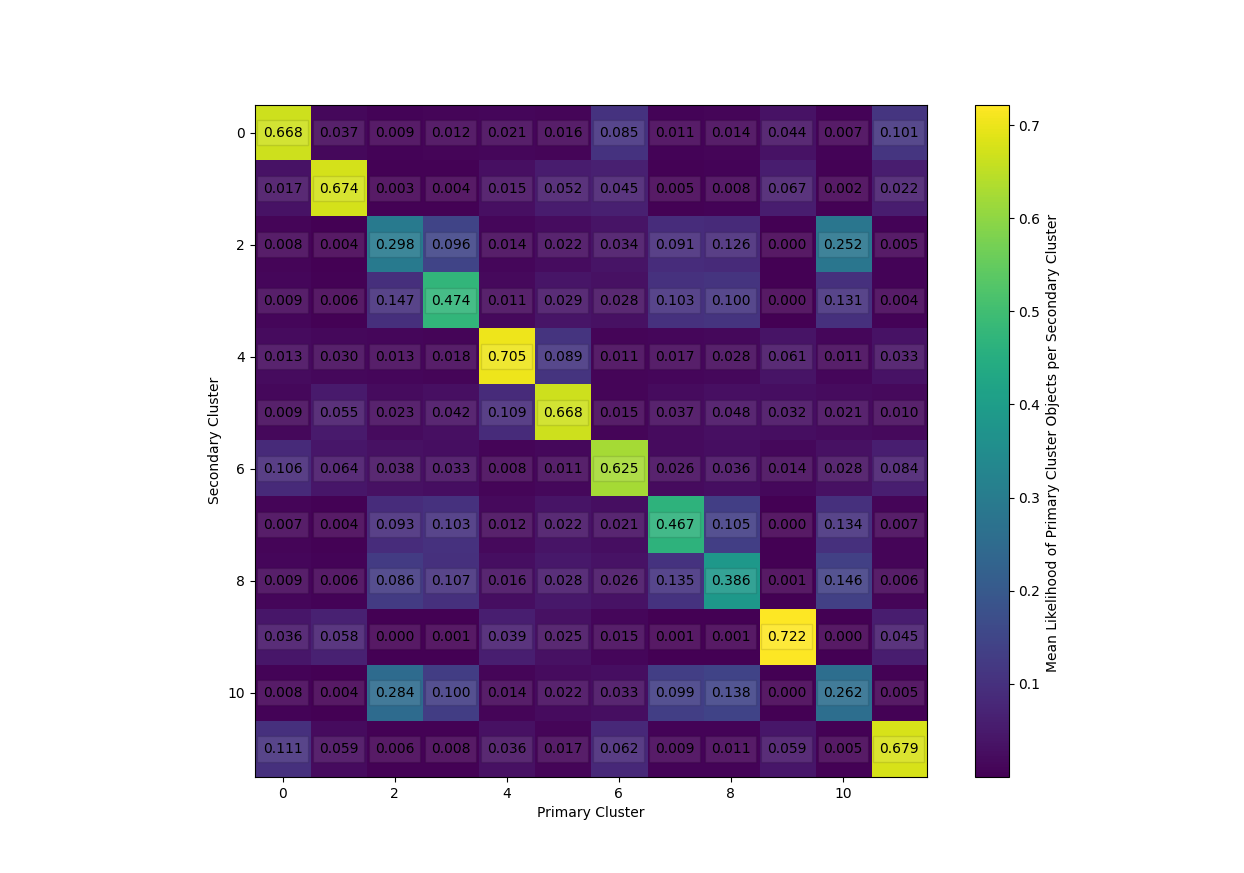}
    \caption{Cluster overlap matrix, which shows the probability that an object in the test set which resides in one cluster could also reside in another cluster. The primary clusters along the {\it x}-axis refers to the assigned label to each galaxy, based on their highest likelihood among the mixture components. The colours and numbers in each element are the mean likelihoods of objects for the secondary mixture component. For example, an object in Cluster $2$ ({\it x}-axis), has an average likelihood of being assigned to Cluster $10$ ({\it y}-axis) of $0.284$.}
    \label{fig:Cluster_Overlap}
\end{figure*}

\begin{figure*}
    \centering
    \includegraphics[width=\textwidth]{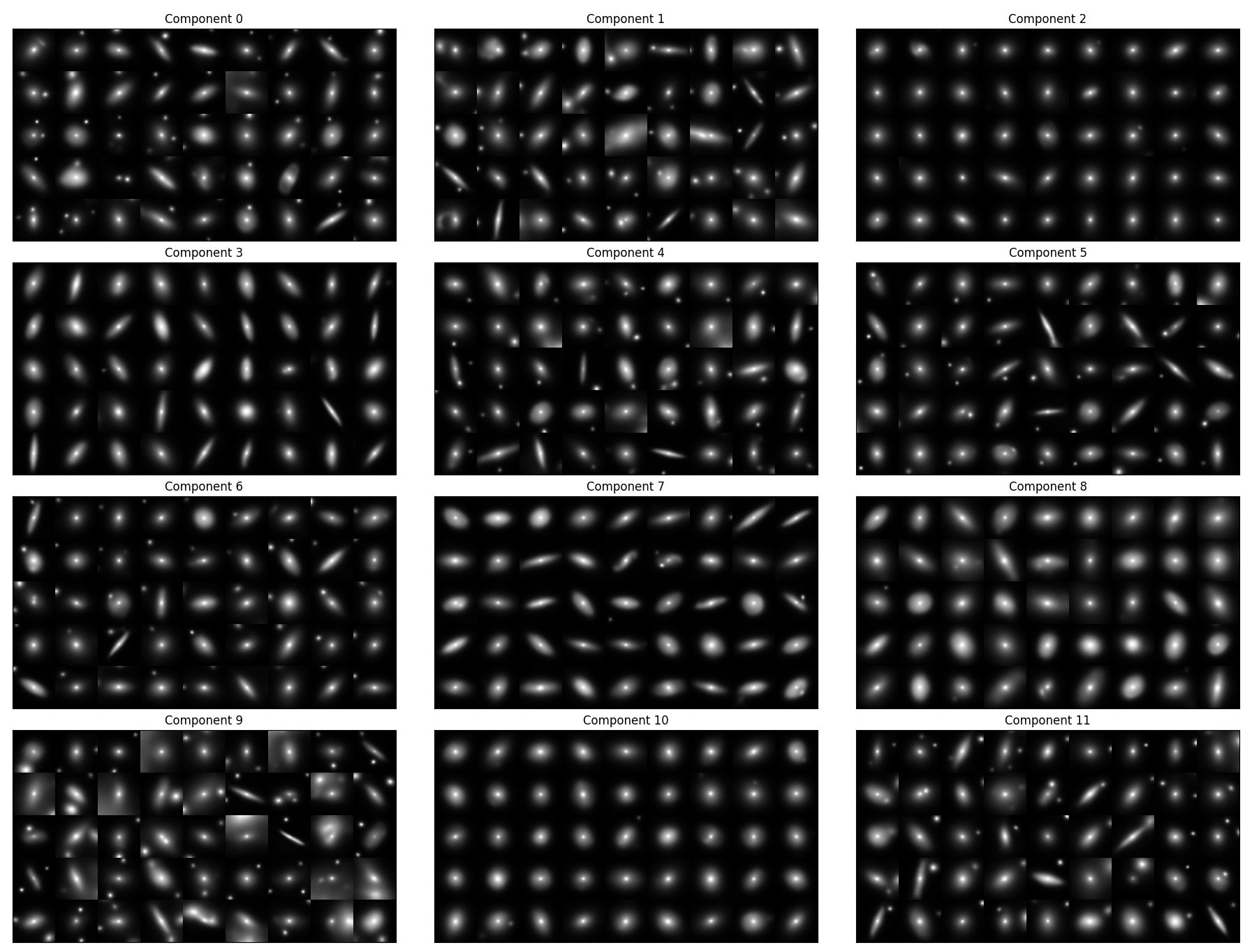}
    \caption{Image reconstructions of test data objects assigned to each cluster. This figure shows a random selection of galaxies from those that have the appropriate cluster label. Images are shown with a linear scale and pixel range of $(0,1)$.}
    \label{fig:Clusters_Recon_RandAssign}
\end{figure*}

\begin{figure*}
    \centering
    \includegraphics[width=\textwidth]{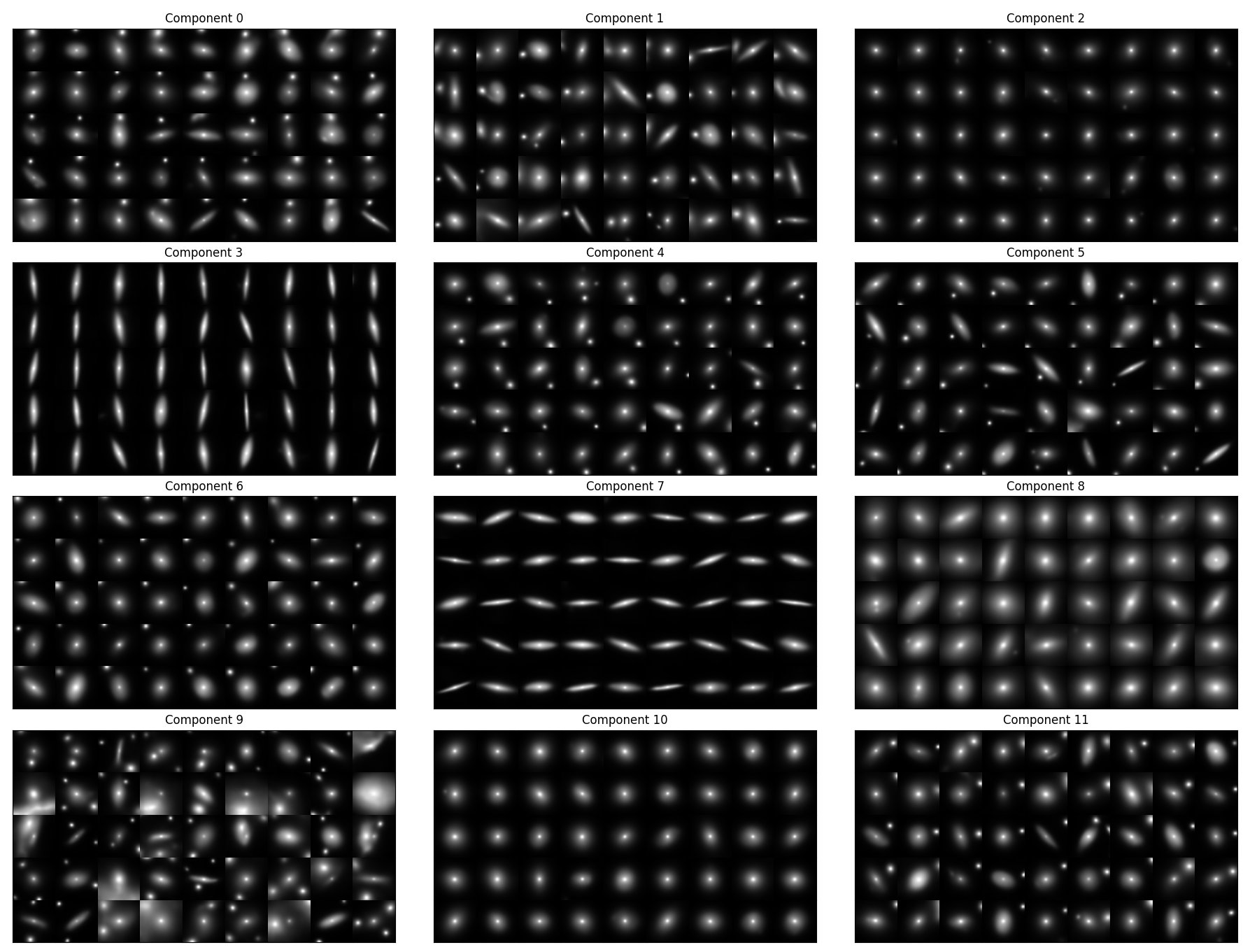}
    \caption{Image reconstructions of test data objects assigned to each cluster. This figure shows the objects with the highest cluster likelihood of objects from those that have the appropriate cluster label. Images are shown with a linear scale and pixel range of $(0,1)$.}
    \label{fig:Assigned_Recon_HighProb}
\end{figure*}

\begin{figure*}
    \centering
    \includegraphics[width=\textwidth]{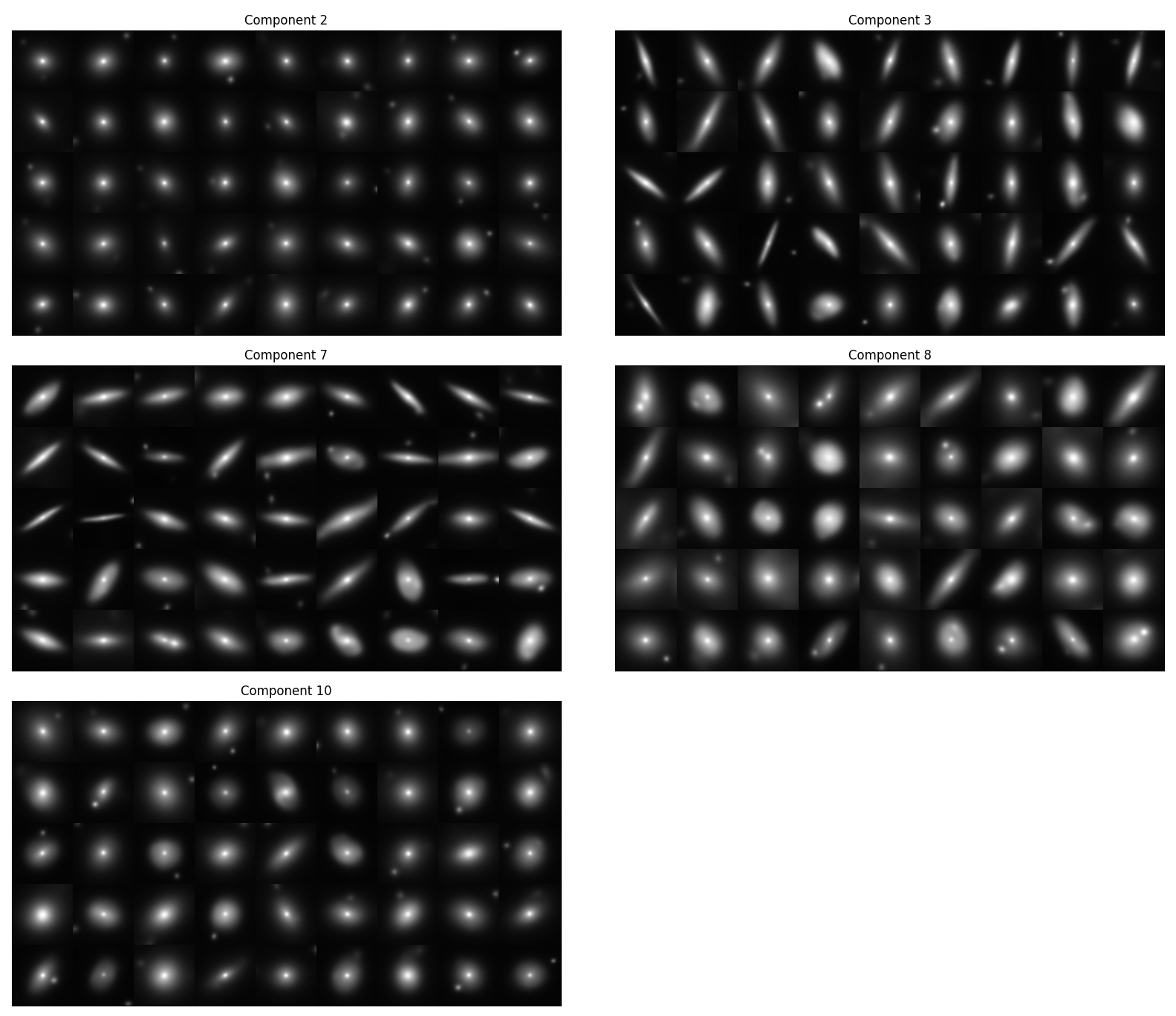}
    \caption{Image reconstructions of test data objects divided by their most likely morphological cluster (see discussion in Section \ref{sec:Clustering}. For each component, we show those galaxies that have the highest cluster likelihoods, selected from those that were not assigned to these components due to the presence of a secondary source. Images are shown with a linear scale and pixel range of $(0,1)$.}
    \label{fig:Secondaries_Recon_HighProb}
\end{figure*}

Before visually assessing the galaxies, let us begin by looking at the GMM properties themselves. In Table \ref{tab:cluster_assignmnts}, we show how many galaxies are assigned each cluster label based on their highest probability component in the GMM. We can see that there is a good range of coverage between the components, with clusters 2, 7 and 3 being the most populated, while cluster 9 is the least populated. Since the mixture model is probabilistic, we can also interrogate the degree of overlap between the clusters in a number of ways. First, we show three different cut off values for $\gamma_c$, the predicted cluster probability for each galaxy, which could be used in a similar manner as cuts in the Galaxy Zoo 2 catalog for the basis of selecting clean samples. Recall that $\gamma_c$ is normalised as per Equation \ref{eqn:minmax}. We also apply a softmax operation to $\gamma_c$ to force the values to sum to unity. With $\gamma_c > 0.1$, a large number of galaxies appear in multiple components, indicating that this threshold is not high enough to make any clean cuts. At $\gamma_c > 0.3$, the cluster assignments drop below the values of those found from the highest probability component for each object. The number of objects that have $\gamma_c > 0.3$ in any component is 28,781, just 70\% of the full test sample. The final cut of $\gamma_c > 0.5$, shows that clusters 2 and 10, which were the most populated, are now empty, while the less populated clusters now retain a higher proportion of their occupants.

This relationship between the cluster occupancies would seem to indicate that the large components have a higher degree of overlap with each other than with the full mixture. There are a number of ways to measure the overlap between clustering results, and we show a few here. Table \ref{tab:cluster_assignmnts} shows the mean Simplified Silhouette Score, $s(i)$, for objects that are assigned each component label. The silhouette score for a sample, $i$ is calculated as:

\begin{equation}
\mathbf{s(i)=}\left\{\begin{array}{ll}
\mathbf{1-a(i) / b(i)}, & \textbf { if } \mathbf{a(i)<b(i)} \\
\mathbf{0}, & \textbf { if } \mathbf{a(i)=b(i)} \\
\mathbf{b(i) / a(i)-1}, & \textbf { if } \mathbf{a(i)>b(i)}
\end{array}\right.
\end{equation}

\noindent where $a(i)$ is the mean distance between $i$ and its fellow cluster members. $b(i)$ is calculated by finding the smallest mean distance between $i$ and the members of each other cluster in the model \citep{ROUSSEEUW198753}. A silhouette score is a measure of how similar an object looks to its fellow cluster members, compared to the samples outside its cluster, where higher scores mean that each cluster is more similar to itself than the full sample. Cluster 2 has the highest mean silhouette score, which indicates that its members are more similar to each other than they are to the full sample, whereas for example, clusters 4, 9 and 10 have lower values, suggesting there are similarities between these objects and the rest of the sample. We note that the mean score for the full sample is $-0.074$, which is close to zero and suggests, as the other measurements do, that there is a high degree similarity between the clusters.

As a final test of the intercluster overlap, we provide Figure \ref{fig:Cluster_Overlap}. This figure shows that for each component along the horizontal axis, the mean $\gamma_c$ for galaxies with that label in each component of the mixture. These mean probabilities show quite clearly the groupings of clusters that appear to overlap on different subsets of objects. Choosing cluster 2 on the horizontal axis, for example, we see that those objects have an almost equal probability of being assigned a cluster 10 label. In contrast, an object in cluster 9 has almost no chance of being in cluster 2. This analysis shows, in several ways, what one might expect for such a model of galaxy morphology: that there are no true `distinct' classes of objects, and that what we describe as a galaxy's morphology is a combination of many different components. While not unique to astronomy, the problem of unbalanced and blended classes of objects in image recognition and unsupervised clustering is particularly prominent with morphological classification, and it has been argued that this makes astronomy an ideal field for pushing boundaries in machine learning research \citep{Kremer_2017}.

We will now turn to visually inspecting the objects within the clusters. Since we know what components ought to contain similar objects, we can use that information to inform us as to why they have been separated. For each component, we show the objects that have the highest probabilities of objects with that label, and a random selection of objects, in Figures \ref{fig:Clusters_Recon_RandAssign} and \ref{fig:Assigned_Recon_HighProb}, respectively. Shown here are the reconstructed images of these objects, however the input images are provided in the Appendix for reference. An immediate, but perhaps surprising, observation can be made from both of these figures: the components broadly fall into two groups, that is objects with and without secondary sources in the images. We also see that for the components that have objects that are highly dependent on orientation, that those objects have been split between components.

We find five clusters that contain galaxies without secondary sources, six that do contain secondaries, and one final cluster that appears to be predominantly corrupted images or very complex systems. Of the five clusters without secondaries, two contain disc galaxies with high inclinations that are separated by rotation, one appears to mainly contain `larger' galaxies such as face-on disks and elliptical galaxies. The final two clusters are difficult to separate visually, based on the reconstructed images alone. The immediate difference appears to again be size and brightness, but there also appears to be a factor of surface brightness profile, as the galaxies in cluster 2 (top right) appear more compact and centrally dominated.

The six clusters containing objects that generally have fewer than two secondary sources pose a significant challenge for AstroVaDEr. These galaxies have not been separated at all based on the morphology of the primary source, but rather on the position of the secondary sources. In the top left cluster, the secondaries are all positioned along the upper most edge of the images, while the centre top objects all have a secondary to their left. While this may seem to be nonsensical in the context of morphological classification, it does tell us something very interesting about AstroVaDEr, which is related to the very nature of autoencoder architectures in general.

The clusters which are morphological dominated are those in Figure \ref{fig:Cluster_Overlap} which have lower likelihoods among their defined members (values along the diagonals in the confusion matrix), clusters $2$, $3$, $7$, $8$, and $10$. This makes sense when we consider that the remaining components contain galaxies of all shapes. A galaxy without a bright secondary source that is an edge-on disk would be equally likely to be in any of the secondary source clusters, but only have a high likelihood in one or two of the morphological groups. Contrast this with an edge-on galaxy that has a bright secondary, which would only fit in the component that matches the position of its companion and the edge-on clusters. We also note that this agrees with the silhouette scores in Table \ref{tab:cluster_assignmnts}, all the components except $2$ and $10$ (top right and bottom centre in Figures \ref{fig:Clusters_Recon_RandAssign} and \ref{fig:Assigned_Recon_HighProb}, respectively) have negative silhouette scores and the lowest values correspond to components that contain galaxies with a range of morphologies. Components $2$ and $10$ have positive scores, and contain objects which could be considered very "average" or "normal", i.e. they are round and regular and smooth looking. Even components $3$, $7$ and $8$ (second row-left, third row-middle and third row-right in the Figures) have scores much closer to $0$ than the groups with secondary sources.

The GMM we trained is a representation of the latent space that objects are embedded into, and the types of objects that are grouped together in the 20D space depends on how much of the network capacity is being used to learn those features. In this iteration, AstroVaDEr has apparently committed a significant fraction of its capacity to accurately recreating the presence, size, brightness and position of the secondary, tertiary, etc. sources in these images. This is an example of the ways in which unsupervised learning algorithms can produce representations of data in ways that humans brains would not consider doing. In Section \ref{sec.Data} we deliberately chose {\it not} to rotate our images as part of the pre-processing before training. In part, this was because we expected that horizontal and vertical flips would impart some degree of rotational invariance, but also because we were interested in the specific properties of the embedded space without those rotations.

We do note, however, that simply including random rotations in the network does not guarantee that it will not still use the same capacity to store the orientation of an object. Consider a supervised CNN where the goal is to reproduce known labels, in a sense the network `knows' that a rotated image should have the same label and so the rotation is managed as part of the learning. Under unsupervised assumptions, however, the network does not know that two rotated images could be the same object, and as such has no pressure to give them the same label\footnote{In fact, rotational invariance may not even be the correct mode of thinking, as it would imply that the network output (i.e.\ the reconstructed images) would be invariant to rotation. Instead, it could be more useful to consider a possible rotation `equivariance', where the assigned cluster does not change when an image is rotated, and subsequently, similar objects which have relative rotations should also be assigned the same labels. For more information on rotational invariance/equivariance, and one potential way this can be addressed, see \cite{2020arXiv200504613P}.}. For AstroVaDEr, when an image is rotated it will naturally end up in a different region of the latent space, and therefore be better represented by a different component. We will explore this in Section \ref{sec:generative}, and demonstrate how rotation is actually embedded within the different latent variables.

One of the benefits of the probabilistic nature of GMMs, and in particular the model trained here which has a high degree of overlap between its components, is that we can choose to ignore certain components. Throughout the clusters that have secondary sources, we noted that the overall morphologies are ignored. However, if we take the galaxies assigned to those clusters, we can then give them a secondary label based on the probabilities of the five morphological clusters. In Figure \ref{fig:Secondaries_Recon_HighProb}, we show the highest probability objects from the 13,000 objects with secondary sources (excluding those in the cluster of corrupted and complex images), in each of the purely morphological clusters. Doing so allows us to retrieve a morphological label for those galaxies, despite the fact that the network primarily assigns them to other components.

These clustering results, while they may not reflect human intuition, are clearly doing a sensible job of providing some systematic measure of a galaxy's morphology. Equally, they give us a greater insight into the inner workings of unsupervised clustering in the context of morphological features, which can direct future work in refining these methods.

\subsection{Generative Properties of AstroVaDEr}
\label{sec:generative}

\begin{figure*}
    \centering
    \includegraphics[width=\textwidth]{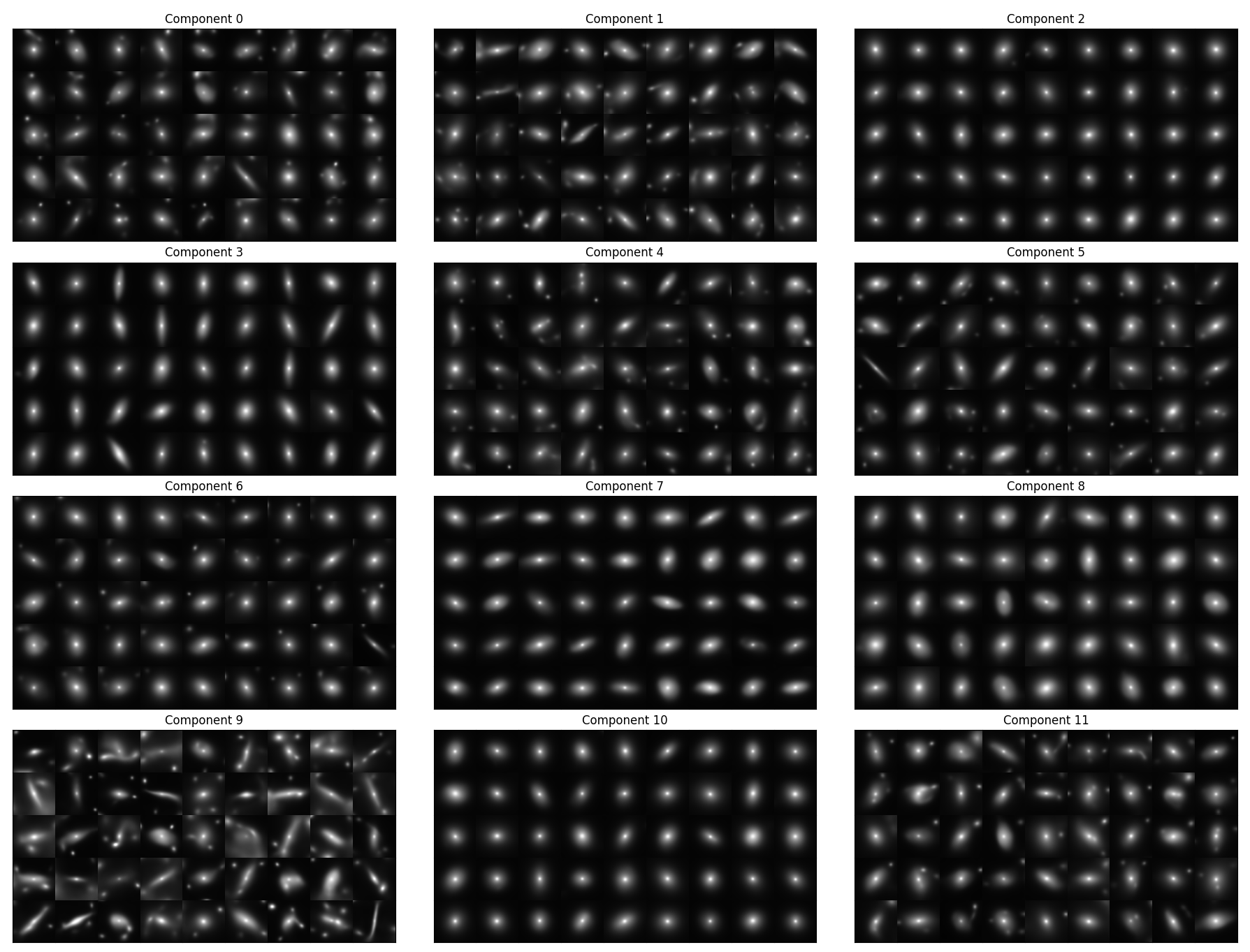}
    \caption{Synthetic images generated using the AstroVaDEr network. For each mixture component, $c$, we randomly sample from the Multivariate Gaussian $\mathcal{N}(\mu_c, \sigma_c^2\boldsymbol{I})$. The sampled vectors are passed into the decoder arm of AstroVaDEr. Images are shown with a linear scale and pixel range of $(0,1)$.}
    \label{fig:Generative_AllClusters}
\end{figure*}

Aside from clustering, the other main task of AstroVaDEr is to act as a generative network. The generative properties allow us to create synthetic images, as we will demonstrate, but also it can be leveraged to allow us to investigate the properties of the latent variables that make up the embedded space. We will begin by simply creating new images and seeing how they hold up compared to the reconstructions and original images. Then we will explore the latent space by interpolating through the space and observing how the visual properties of the generated images change.

One of the ways the generative model can be used in practice is in the data-driven production of a realistic synthetic imaging data set that can be used to test data analysis pipelines intended for next generation surveys, before those surveys beginning releasing public science data to the community. Synthetic image generation with AstroVaDEr works as follows:

\begin{enumerate}
    \item Choose a cluster, $c$, either by choice or from $c \sim \operatorname{Cat}(\boldsymbol{\pi})$,
    \item Sample $\boldsymbol{z}$ from the Multivariate Gaussian $\mathcal{N}(\mu_c, \sigma_c^2\boldsymbol{I})$,
    \item Feed $\boldsymbol{z}$ in the decoder network to generate a synthetic $\boldsymbol{\hat{x}}$,
    \item Repeat until the desired number of images have been generated.
\end{enumerate}

\begin{figure*}
    \centering
    \includegraphics[width=1\textwidth,angle=0,origin=c]{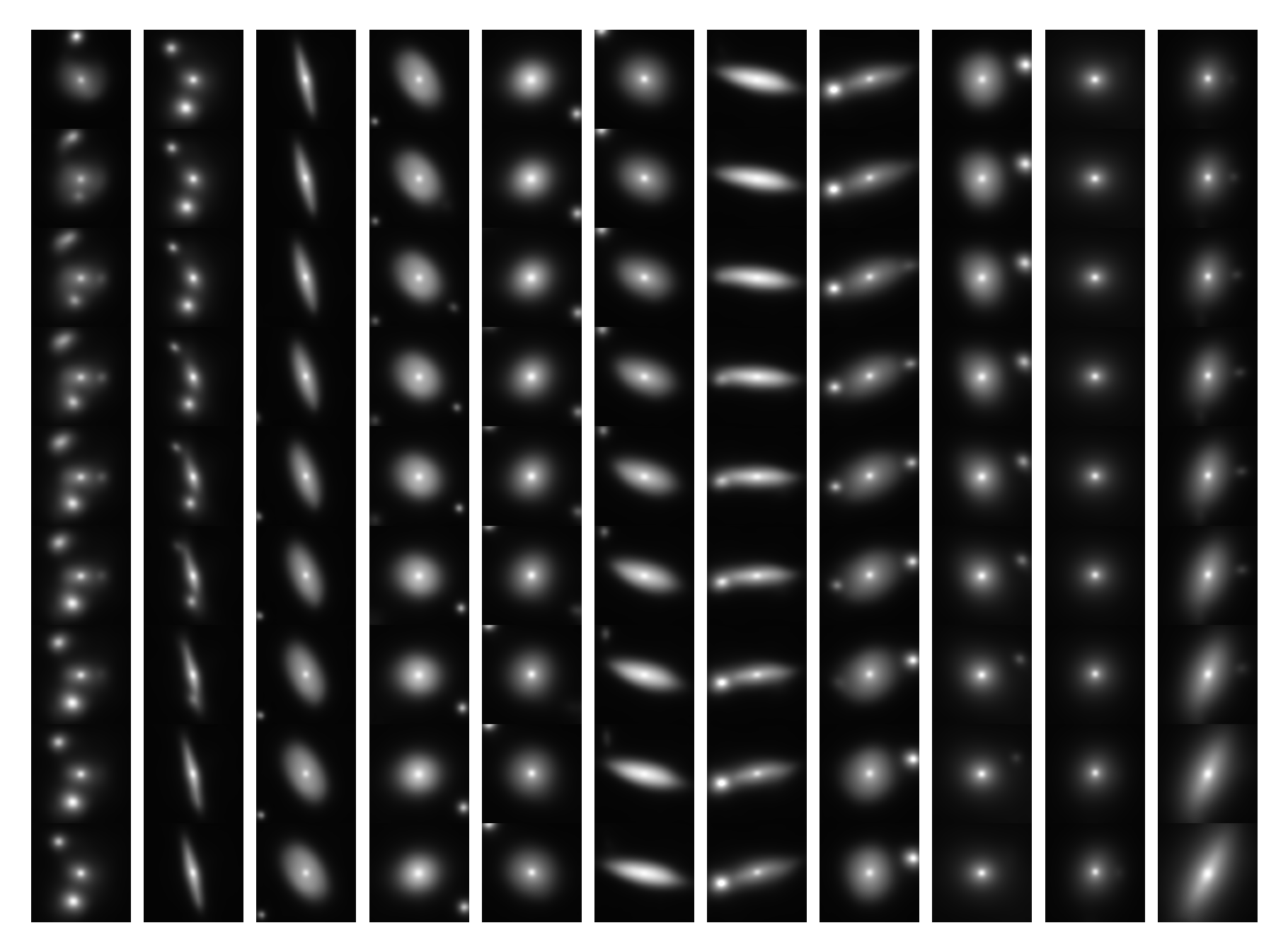}
    \caption{Synthetic images generated using the AstroVaDEr network. For each component, we select the galaxy with the highest cluster likelihood, and then order the clusters based on the shorted route through the latent space with a travelling salesperson algorithm and interpolate along that path. The first and last image of each column are reconstructions of real objects, and the objects between are synthetic. The sampled vectors are passed into the decoder arm of AstroVaDEr. Images are shown within a pixel range of $(0,1)$.}
    \label{fig:Generative_ClusterInterp}
\end{figure*}

Note that we have a choice in whether to generate a fixed number of images for a particular cluster, or sample over the full mixture and generate a data set that reflects the morphological distribution of the training data. For example, we may choose to only sample from the five morphological clusters, as we know that primary properties of those components are related to key morphological characteristics. To begin with, let us recreate Figure \ref{fig:Clusters_Recon_RandAssign} by randomly sampling from each component separately. The results of this generative sampling are shown in Figure \ref{fig:Generative_AllClusters}. For the five morphological clusters in question, the resemblance to the image reconstructions is striking. Each of the clusters contains objects that look very similar to the random selection show in Figure \ref{fig:Clusters_Recon_RandAssign}, and a few even resemble the highly probable objects in Figure \ref{fig:Assigned_Recon_HighProb}.

There appears to be a larger range in the quality of the synthetic images in the remaining components of the model. Individually, most of the synthetic images look quite plausible and exhibit morphologies that are interesting and complex, but not outside the realms of reality. Comparing the synthetics in this Figure to the reconstructions in Figures \ref{fig:Clusters_Recon_RandAssign} and \ref{fig:Assigned_Recon_HighProb}, however, it does seem like AstroVaDEr is producing more objects with these complex morphologies. While the relative positioning of the secondary sources is mostly consistent, the fact that these clusters contain galaxies with all morphologies leads to some blending of different characteristic properties that definitely limits their quantitative value. Some images take on a `wispy' like quality, where multiple secondary sources have blended together, almost resembling (but certainly not a representation of) gravitational lenses. The bottom left cluster, which is made up of anomalous, corrupted and complex images in the training data, contains some very strange objects: some resemble merger remnants, but interpretations of these galaxies may be more at home in a Salvador Dali exhibit than in a synthetic imaging catalogue.

While it is clear that the generative process in AstroVaDEr is by no means perfect, performing these experiments gives us valuable insights into how the model may be improved. The final set of experiments we perform on the generative properties of the network relate to understanding how the network makes the decisions it does, by digging into the structure of the embedded space. Using the generative properties, we can explore the latent variables and produce images that demonstrate what features are being learned.

Let us first consider the 12 learned clusters. What happens to the generated images as we move through the latent space between each cluster? We select the highest probability object for each cluster, and then calculate the shortest route between each object within the embedded space using a travelling salesperson algorithm. Using that route, we linearly interpolate new latent vectors and generate images. Figure \ref{fig:Generative_ClusterInterp} shows the results of this generative sample. As we expect, transitions between clusters are smooth, indicating that there are no discontinuities in the latent space where unrealistic objects might be formed in the decoding process. We can gain a sense here of how features such as the axis ratio and orientation vary through the space, and can see how secondary sources can be generated with varying intensities and positions.

\begin{figure*}
    \centering
    \includegraphics[width=1\textwidth]{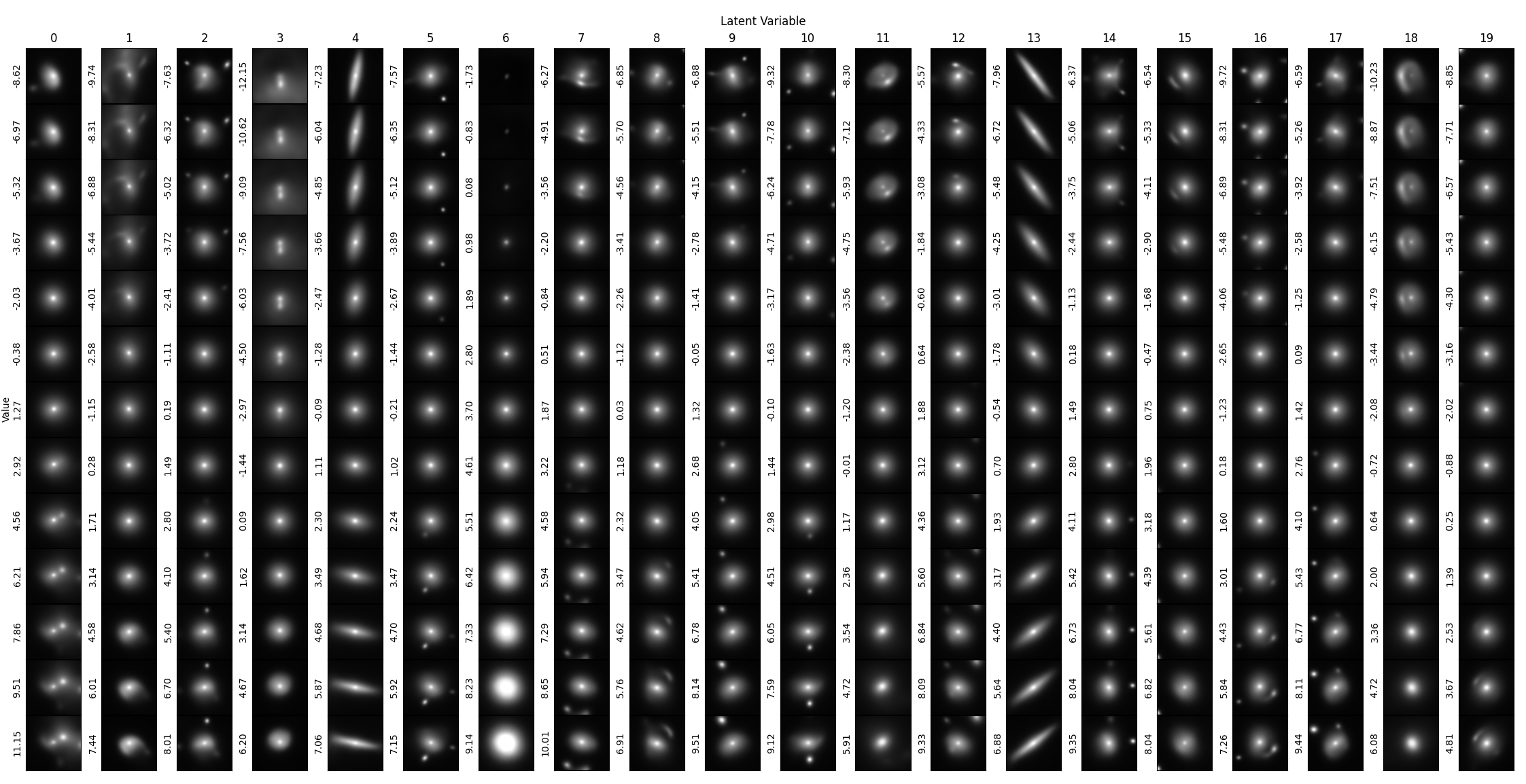}
    \caption{Synthetic images generated using the AstroVaDEr network. We calculate the mean, maximum and minimum values of each latent variable, and in each column we linearly interpolate one variable between the minimum and maximum, while keeping all other variables fixed. The interpolated vectors are passed into the decoder to generate synthetic images. Images are shown within a pixel range of $(0,1)$.}
    \label{fig:Generative_LatentMinMax}
\end{figure*}

It is clear from the synthetic images that the latent space is quite complex, and with just 20 latent variables it is difficult to disentangle various features. The most obvious is the secondary sources: the network needs to be able to control the number of sources, their intensity (which it seems to do independently in some cases), and their positions. In Figure \ref{fig:Generative_LatentMinMax}, we explore each latent variable individually by generating latent vectors which vary between the minimum and maximum of each variable. We keep the other variables fixed at their mean value in this process. Within each row of the figure, we can see the varied morphological features it is controlling, compared with the `mean galaxy' down the centre column. For example, we see that the galaxy's brightness is almost entirely controlled by a single variable (column $7$), and the axis ratio and orientation appears to be split over two variables. As expected from our previous discussions, we can see quite clearly that much of the network capacity is given to controlling the positions, intensity and quantity of secondary sources. Minimising the network footprint of these secondary sources appears to be a key challenge in improving the generative and clustering capability of the network.

\begin{figure*}
    \centering
    \includegraphics[width=1\textwidth, clip, trim=50mm 90mm 50mm 90mm]{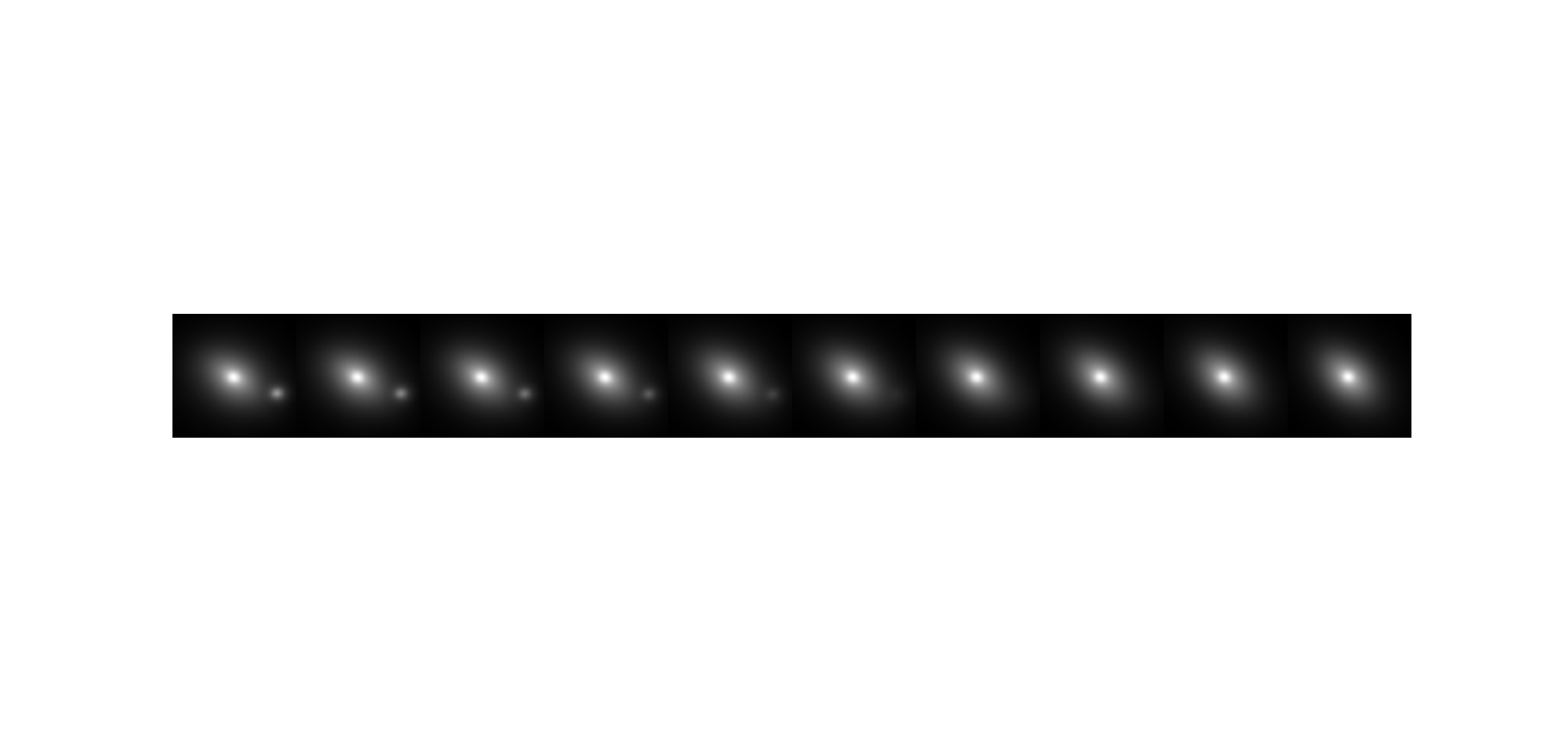}
    \caption{Example of deblending a target galaxy with the secondary sources in the image. The left and right most images are reconstructions of each galaxy, and the intermediate synthetic galaxies are generated by interpolating between the latent vectors of the two objects.}
    \label{fig:Generative_Deblend}
\end{figure*}

One positive aspect of this experiment is that we can see that none of the latent variables are doing `nothing,' per se. All of the latent variables have a role to play in embedding the images. If anything, the most crucial take away here is that there is space for more latent variables to be added. The question becomes `how many variables is too many?'. What we do not want to see happen is, say, doubling the number of features leading to an improvement in reconstruction loss, but a collapse of the clustering optimisations.

One final point of note from Figure \ref{fig:Generative_LatentMinMax} is that it demonstrates a potential application of AstroVaDEr (or, potentially better suited, a non-clustering VAE implementation). Scanning across some latent vectors reveals that there is some degree of deblending happening. It may be possible to tailor a version of the model to remove the background/foreground objects from the target samples. Doing this effectively will require a much higher level of disentanglement of the latent variables than AstroVaDEr currently possesses, however, a crude way of doing this can be achieved using the cluster assignments. We first pick a galaxy assigned to one of the clusters that features secondary sources, we then find the pure morphological cluster it relates to, and find the a galaxy that looks similar. We choose the similar looking galaxy by calculating the mean squared error between the object we want to deblend and all the objects assigned to its morphological class. Finally, we interpolate between the two chosen galaxies. The results of this test are shown in Figure \ref{fig:Generative_Deblend}, the left-most and right-most images are reconstructions of real galaxies, and those between are linear interpolations through the latent space.

\begin{figure}
    \centering
    \includegraphics[width=\columnwidth, trim = 0mm 0mm 0mm 0mm, clip ]{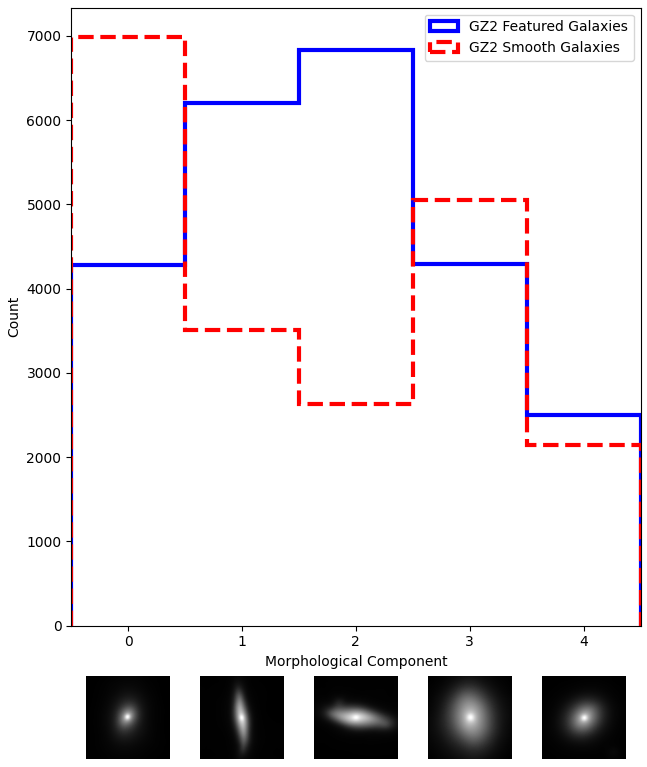}
    \caption{Histogram showing the number of Galaxy Zoo 2 `Featured' (blue solid line) and `Smooth' (red dashed line) galaxies in the $5$ morphological components in the learned GMM. Below each bin in the histogram we show an example reconstructed image of an object from that component.}
    \label{fig:GZ2_Featured_Smooth}
\end{figure}

\begin{figure}
    \centering
    \includegraphics[width=\columnwidth]{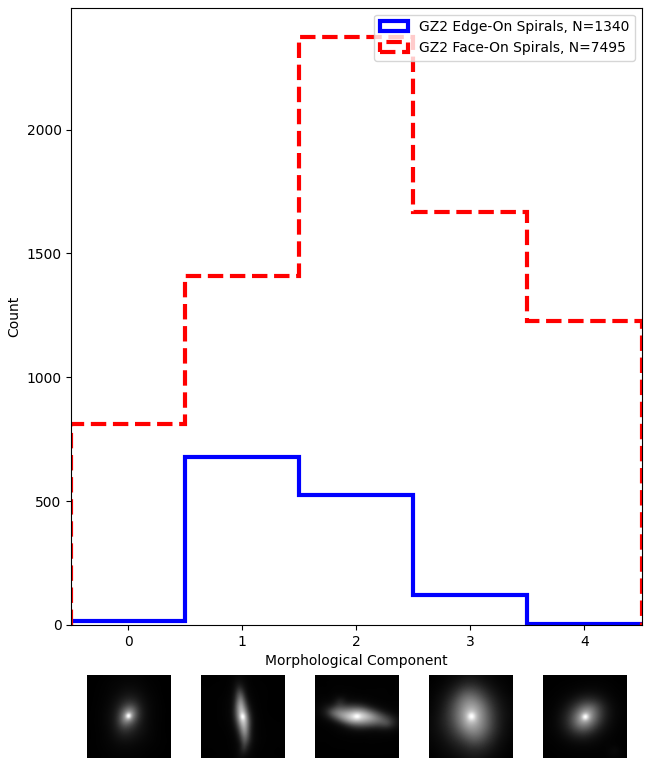}    \caption{Histogram showing the number of Galaxy Zoo 2 `Edge-on' (blue solid line) and `Face-on' (red dashed line) galaxies in the $5$ morphological components in the learned GMM. Below each bin in the histogram we show an example reconstructed image of an object from that component.}
    \label{fig:GZ2_Edge_NoEdge}
\end{figure}

\subsection{Comparing AstroVaDEr Clusters and Galaxy Zoo 2 Classifications}

Finally, we will now consider how AstroVaDEr and its cluster assignments compare with the human classifications collected in GZ2. Given AstroVaDEr's current reconstruction ability, we will concern ourselves with the top level questions in the Galaxy Zoo 2 decision tree: Smooth versus Featured, and Edge-on versus Face-on. As we have identified that only $5$ of the clusters relate to morphology, we restrict this analysis to those clusters only, assigning a label to each test sample galaxy based on the highest probability for each of those clusters (as in \ref{sec:Clustering}). For each cluster we calculate the number of test sample galaxies that lie above a particular cut in the debiased vote fractions for each response. For smooth and featured galaxies we set this cut at $p=0.43$, and for edge-on and face-on galaxies we apply the featured cut and then use $p=0.715$ to select clean samples of each. These values are those recommended in \cite{2013MNRAS.435.2835W} for clean samples.

In Figure \ref{fig:GZ2_Featured_Smooth} we show the number of galaxies that are smooth or featured based on the GZ2 debiased vote fractions. Morphological component $0$ in this Figure has a higher number of smooth galaxies, implying that this cluster of objects is mainly populated by early-type galaxies. Components $1$ and $2$, the components we identified as being mainly high axis ratio galaxies, have a higher number of featured objects, which is to be expected. Finally, components $3$ and $4$ have more balanced numbers of smooth and featured galaxies, which we infer as meaning that the primary split here is likely on size and brightness profiles.

In Figure \ref{fig:GZ2_Edge_NoEdge} we show the distribution of featured galaxies which are voted as "Edge-On" and "Not Edge-On" (i.e.\ face-on) by GZ2 users. Remember that the chosen cut is intended to produce `clean' samples, so galaxies that are unclear in whether they are edge-on or not are not included. We first note that there are many more face-on galaxies than edge-on galaxies, but this is to be expected as only fairly extreme axis-ratios are consistently voted as edge-on in the GZ2 scheme. Components $0$, $3$, and $4$ in this Figure, as expected, have almost no edge-on objects between them. given that these clusters pick up smooth and featured galaxies, it makes sense that the late-type galaxies in this group would need to be face-on in order to "look" the same to the network.

The most curious splits in this comparison are in clusters $1$ and $2$, which we expected to be mainly edge-on type galaxies. In component $1$, the number of edge-on objects is roughly half that of the face-on objects, but in component $3$ there are $4$ times as many face-on galaxies as there are edge-on galaxies. This brings into question how these two clusters are really behaving, as we initially believed that rotation was the main discerning feature of these groups. It appears from this analysis, that these clusters also discern between axis ratios as a secondary feature. We have discussed throughout this work the importance of latent disentanglement, as this is a clear example of how latent features can be entangled in the clustering paradigm.

It is clear that while AstroVaDEr does not, and indeed is not designed to, reproduce the Galaxy Zoo classification scheme, it is certainly interesting how the morphological groups we have identified compare with the independent user generated labels. In the future we hope to improve the reconstruction and disentanglement of AstroVaDEr's clustering to find finer grained features in which the galaxies are grouped together. In particular, trying to pick out distinct morphological features such as bars and rings, and possibly gravitational lenses and low surface brightness features in next generation surveys that have sufficient resolution and depth. We also note that this comparison with Galaxy Zoo brings the exciting possibility of combining unsupervised learning and citizen science, wherein we envision a possible platform where citizen scientists could be employed to find the common features between objects in unsupervised clusters.

\section{Future Work}
\label{sec:future_work}

The architecture presented in this work was chosen for demonstrative purposes, and by no means fully optimised to the task of galaxy classification. In this section we shall briefly discuss how AstroVaDEr will be improved and optimised in the future for general release and application to next-generation astronomical surveys.

The main point to address in improving AstroVaDEr for scientific applications is the balance between the reconstruction loss and the clustering loss. There is much room to improve both the quality of the image generation (both synthetic and reconstructed outputs) and the latent variable/cluster disentanglement. Much work has been done to investigate the sparsity problem in Variational Autoencoders, such as Balance VAE \citep{2019arXiv190305789D,2020arXiv200207514A} and Regularised Autoencoders \citep{2019arXiv190312436G}. These models are designed to work with unit Gaussian priors, but it may be possible to develop a model under these schemes that works with a GMM. It may also be possible to develop a clustering version of Introspective VAE \citep{2020arXiv200512573K}, which utilises a GAN-like structure where the decoder output is re-embedded and trained antagonistically with the encoder.

AstroVaDEr is currently coded in an outdated version of \texttt{Keras} which still relies on \texttt{Tensorflow} Version 1.15 \citep{Tensorflow2015-whitepaper}. Further development of the platform will involve updating the model to run in \texttt{Tensorflow} 2.0. Among other updates to the \texttt{Tensorflow} API, we hope to include the \texttt{Tensorflow} Probability\footnote{\texttt{Tensorflow} Probability: \url{ https://github.com/tensorflow/probability}} module, which includes optimised probability distributions for using within machine learning architectures. We will also investigate the inclusion of more complex convolutional blocks, such as ResNet blocks \citep{he2015deep}. Finally, we note that the order of the up-sampling and convolutional layers in the decoder is something not universally agreed upon in CAE architecture, so we will test using convolutional blocks before or after upsampling the decoded activity maps. We also plan on changing our training catalog to imaging from the Hyper-Supreme Cam Subaru Strategic Program \citep{2018PASJ...70S...4A}, in an effort to tailor the network toward future Vera Rubin Observatory operations.

\section{Conclusions}
\label{sec:conclusions}

We have presented here a demonstration of AstroVaDEr, an Astronomical Variational Deep Embedder. This network has been developed to perform unsupervised clustering of images of galaxies imaged in the Sloan Digital Sky Survey and classified by citizen scientists in Galaxy Zoo 2. We provide a comprehensive overview of the theoretical background of Variational Autoencoders and Variational Deep Embedding, and show how we implement cutting-edge optimisations of these networks within our model.

We train AstroVaDEr on around $160,000$ images of nearby galaxies to embed the images into 20 latent variables which have Gaussian distributions, reconstruct the images from the latent embedding, and finally to cluster the images within the later space using a Gaussian Mixture Model (GMM). The trained network is able to produce cluster labels and probabilities for new images of galaxies, as demonstrated with a test sample of approximately $41,000$ objects, and also to generate synthetic images of galaxies randomly drawn from the GMM.

Reconstructed and synthetic images produced by AstroVaDEr show qualitatively realistic properties in terms of shape, light profiles, bulge presence, axis ratio and size. Currently, we are not able to fully reconstruct finer grained details like spiral arms and galactic bars, and output images tend to have a `blurred' or `smoothed' look. This blurring is mainly due to underlying problems in VAE architecture, which originates in the competition between reconstruction loss and clustering loss. We discuss in Section \ref{sec:future_work} some of the ways this may be addressed. We show how the generative process used in the network is capable of construction a continuous generative space between the different morphological clusters.

The resulting clustering model is able to identify the presence of secondary sources within the images, and also provides a number of morphological groups. We find that galaxies are grouped together based on size, surface brightness distribution, axis ratio and rotation. We compare our clustering scheme to the debiased vote fractions from Galaxy Zoo 2 and find correlations between the unsupervised clusters and `smooth' and `featured' galaxies, and with `edge-on' and `not edge-on' galaxies.

AstroVaDEr has potential to be used with next generation sky surveys as a science-enabling platform. We envision that it could be used to generate large scale synthetic imaging datasets to use in testing and developing data analysis pipelines in preparation for future data releases, for example on three NVIDIA Tesla V100 GPUs we can generate $300,000$ images at $128\times128$ pixels in one minute. At this speed we could generate one hundred million images in less than six hours. For the clustering tasks, we predict that it would take about 7 hours to calculate cluster assignments and probabilities of the one hundred million objects using our current hardware. We also show that, even without improvements to the disentanglement of latent variables, AstroVaDEr demonstrates some capability in deblending primary and secondary sources in the input images.

Development of the network continues, with planned improvements focusing on reconstruction/synthetic image quality and disentanglement of latent variables and clusters. We plan on building AstroVaDEr into a flexible platform that can be used by researchers in a variety of fields, in and out of the astronomical community, with only minimal prior knowledge of machine learning networks. The code used to train the model and produce the results of this paper are available at \url{https://github.com/AshleySpindler/AstroVaDEr-Public}.

\section*{Acknowledgements}

A.S.\ is supported by an STFC Innovation Fellowship (ST/R005265/1). J.E.G.\ is supported by a Royal Society University Research Fellowship (URF/R/180014). This research has made use of the University of Hertfordshire high-performance computing facility (\url{https://uhhpc.herts.ac.uk}). 

The authors thank Christopher Lovell (University of Hertfordshire) and Sandor Kruk (ESA, ESTEC) for their valuable insight. The authors would also like to thank the referee for their kind comments and helpful feedback.

%%%%%%%%%%%%%%%%%%%%%%%%%%%%%%%%%%%%%%%%%%%%%%%%%%

%%%%%%%%%%%%%%%%%%%% REFERENCES %%%%%%%%%%%%%%%%%%

% The best way to enter references is to use BibTeX:

\bibliographystyle{mnras}
\bibliography{AstroVaderBib.bib} % if your bibtex file is called example.bib

% Alternatively you could enter them by hand, like this:
% This method is tedious and prone to error if you have lots of references
%\begin{thebibliography}{99}
%\bibitem[\protect\citeauthoryear{Author}{2012}]{Author2012}
%Author A.~N., 2013, Journal of Improbable Astronomy, 1, 1
%\bibitem[\protect\citeauthoryear{Others}{2013}]{Others2013}
%Others S., 2012, Journal of Interesting Stuff, 17, 198
%\end{thebibliography}

\section*{Data Availability}

The data and code used to produce the training, validation and testing datasets, and the results presented in this paper, are archived on Zenodo at \url{https://doi.org/10.5281/zenodo.4034802}.

%%%%%%%%%%%%%%%%%%%%%%%%%%%%%%%%%%%%%%%%%%%%%%%%%%

%%%%%%%%%%%%%%%%% APPENDICES %%%%%%%%%%%%%%%%%%%%%

\appendix
\renewcommand\thefigure{A\arabic{figure}}    

\begin{figure*}
    \centering
    \includegraphics[width=\textwidth]{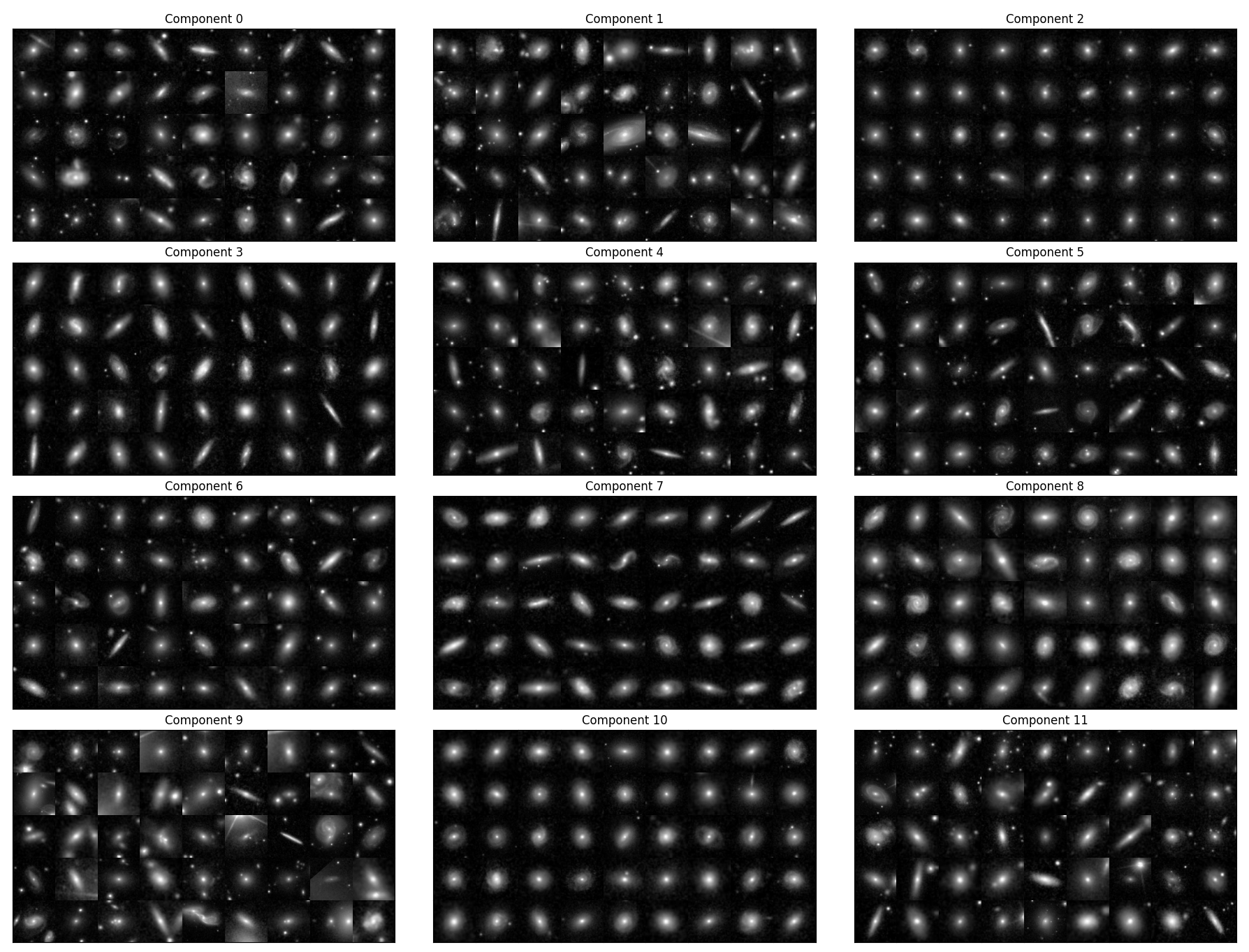}
    \caption{Input images of test data objects assigned to each cluster. This figure shows a random selection of galaxies from among those that have the appropriate cluster label. Images are shown on a linear scale with a pixel range of $(0,1)$.}
    \label{fig:Clusters_Truth_RandAssign}
\end{figure*}

\begin{figure*}
    \centering
    \includegraphics[width=\textwidth]{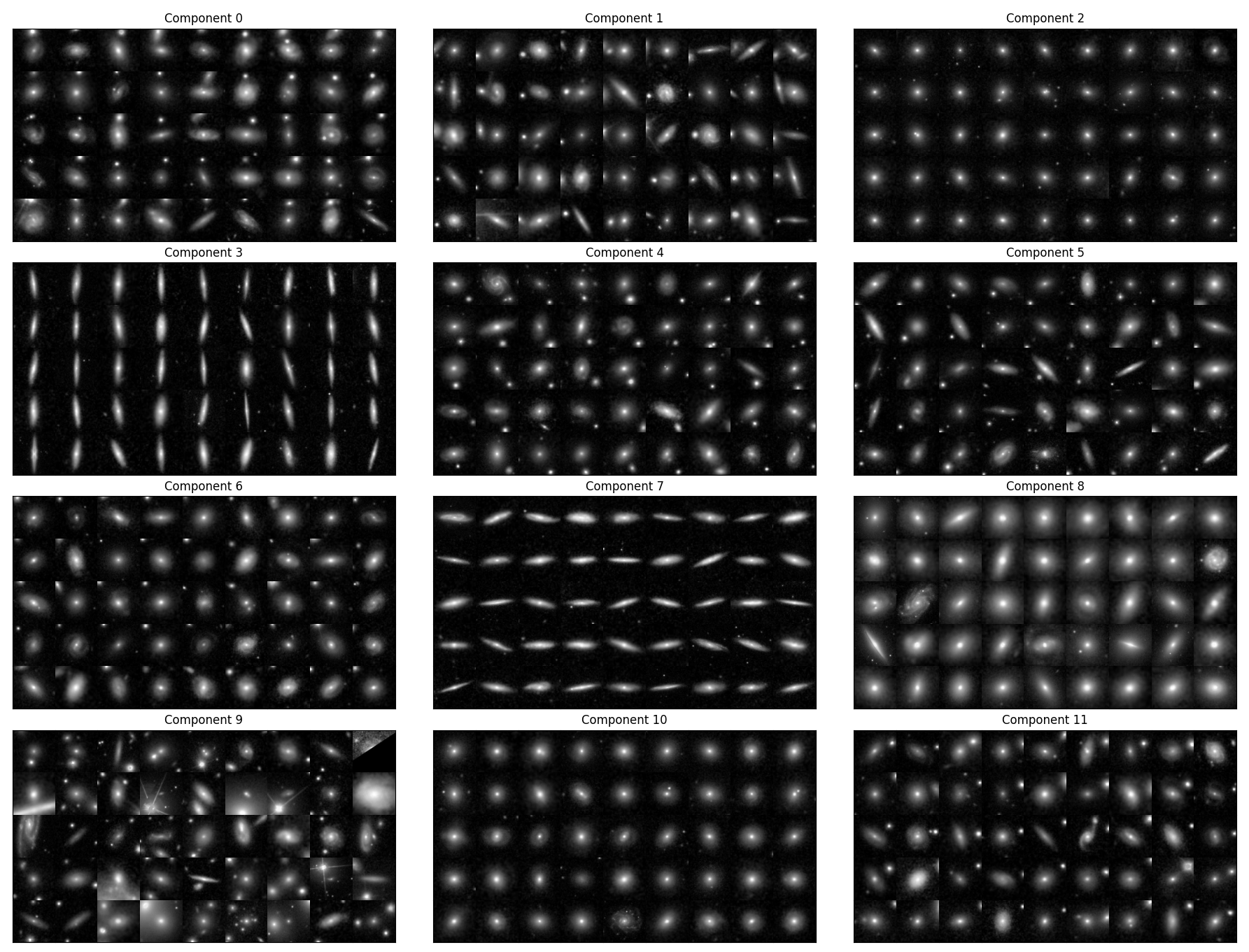}
    \caption{Input images of test data objects assigned to each cluster. This figure shows the objects with the highest cluster likelihood of objects from among those that have the appropriate cluster label. Images are shown on a linear scale with a pixel range of $(0,1)$.}
    \label{fig:Assigned_Truth_HighProb}
\end{figure*}

\section{Ground Truth Images of Clustered Galaxies}

For comparison purposes, we provide the original input images of galaxies included in Figures \ref{fig:Clusters_Recon_RandAssign} and \ref{fig:Assigned_Recon_HighProb} in Figures \ref{fig:Clusters_Truth_RandAssign} and \ref{fig:Assigned_Truth_HighProb}.

%%%%%%%%%%%%%%%%%%%%%%%%%%%%%%%%%%%%%%%%%%%%%%%%%%

% Don't change these lines
\bsp	% typesetting comment
\label{lastpage}
\end{document}